\documentclass[12pt,a4paper]{article}

\usepackage{graphics}
\usepackage{cite,a4wide}
\newcommand{\beq}{\begin{equation}}
\newcommand{\eeq}{\end{equation}}
\newcommand{\bea}{\begin{eqnarray}}
\newcommand{\eea}{\end{eqnarray}}
\renewcommand{\d}{\delta}

\renewcommand{\L}{\Lambda}
\renewcommand{\b}{\beta}

\renewcommand{\ni}{\noindent}
\newcommand{\n}{\nu}
\newcommand{\m}{\mu}

\newcommand{\s}{\sigma}

\newcommand{\vx}{\vec{x}}

\renewcommand{\P}{{\cal P}}
\renewcommand{\th}{\theta}
\renewcommand{\to}{\overline{\theta}}
\newcommand{\tm}{\theta^M}
\newcommand{\tp}{\theta^{ph}}
\newcommand{\tpo}{\overline{\theta}^{ph}}
\newcommand{\tpom}{\overline{\theta}^{ph}(\tm)}
\newcommand{\intpi}{\int^\pi_{-\pi}}

\newcommand{\oh}{\frac{1}{2}}

\newcommand{\dg}{\dagger}
\newcommand{\non}{\nonumber}

\newcommand{\rf}[1]{(\ref{#1})}
\newcommand{\ra}{\rightarrow}
\newcommand{\pa}{\partial}

\newcommand{\rra}{\right\rangle}
\newcommand{\lla}{\left\langle}
\newcommand{\tr}{{\rm Tr}\,}

\begin{document}

\hfill July 1999

\begin{center}

\vspace{32pt}

  { \bf \Large Vortex Structure vs.\ Monopole Dominance \\
\bigskip
               in Abelian-Projected Gauge Theory }

\end{center}

\vspace{18pt}

\begin{center}
{\sl J. Ambj{\o}rn${}^a$, J. Giedt${}^b$, and J. Greensite${}^{ac}$}

\end{center}

\vspace{18pt}

\begin{tabbing}

{}~~~~~~~~~~~~~~~~~~~~~~~\= blah  \kill
\> ${}^a$ The Niels Bohr Institute, Blegdamsvej 17, \\
\> ~~DK-2100 Copenhagen \O, Denmark. \\
\> ~~E-Mail: {\tt ambjorn@nbi.dk, greensit@alf.nbi.dk} \\
\\
\> ${}^b$ Physics Dept., University of California at Berkeley, \\
\> ~~Berkeley, CA 94720 USA.  \\
\> ~~E-mail: {\tt giedt@socrates.berkeley.edu} \\
\\
\> ${}^c$ Physics and Astronomy Dept., San Francisco State Univ., \\
\> ~~San Francisco, CA 94117 USA. \\
\> ~~E-mail {\tt greensit@stars.sfsu.edu}

\end{tabbing}

\vspace{18pt}

\begin{center}

{\bf Abstract}

\end{center}

\vspace{24pt}

   We find that Polyakov lines, computed in abelian-projected SU(2)
lattice gauge theory in the confined phase, have finite expectation values 
for lines corresponding to two units of the abelian electric charge.
This means that the abelian-projected lattice has at most $Z_2$, rather
than U(1), global symmetry.  We also find a severe breakdown of the 
monopole dominance approximation, as well as positivity, 
in this charge-2 case. These results imply that the abelian-projected 
lattice is not adequately represented by a monopole Coulomb gas; 
the data is, however, consistent with a center vortex structure.  Further 
evidence is provided, in lattice Monte Carlo simulations, for collimation 
of confining color-magnetic flux into vortices.

\vfill

\newpage

\setcounter{equation}{0}
\section{Introduction}

   The center vortex theory \cite{cvt} and the 
monopole/abelian-projection theory \cite{abt}
are two leading contenders for the title of quark confinement mechanism.
Both proposals have by now accumulated a fair amount of numerical support.
To decide between them, it is important to pinpoint areas where the
two theories make different, testable predictions. 
In this article we would like to report on some preliminary 
efforts in that direction.

\begin{figure}[h!]
\centerline{\scalebox{0.75}{\includegraphics{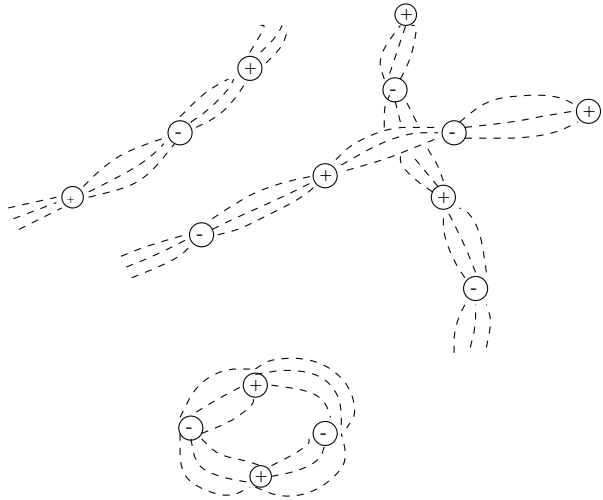}}}
\caption{Hypothetical collimation of monopole/antimonopole
flux into center vortex tubes on the abelian-projected lattice.}
\label{avort5}
\end{figure}

\begin{figure}[t]
\centerline{\scalebox{0.5}{\includegraphics{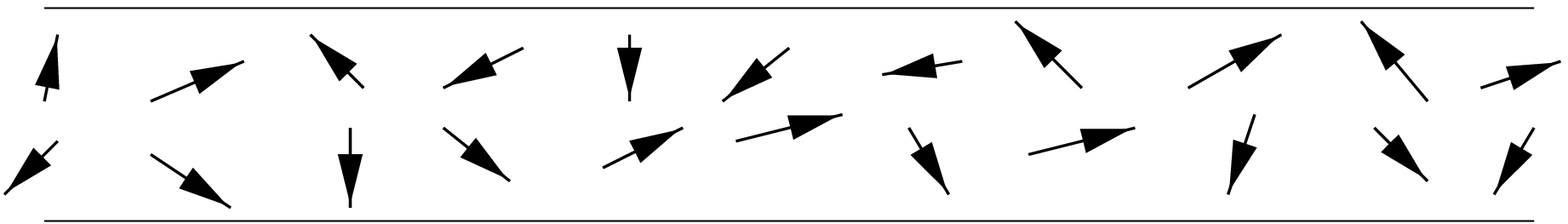}}}
\caption{Vortex field strength before gauge fixing.  The arrows
indicate direction in color space.}
\label{avort1}
\bigskip
\bigskip
\bigskip
\centerline{\scalebox{0.5}{\includegraphics{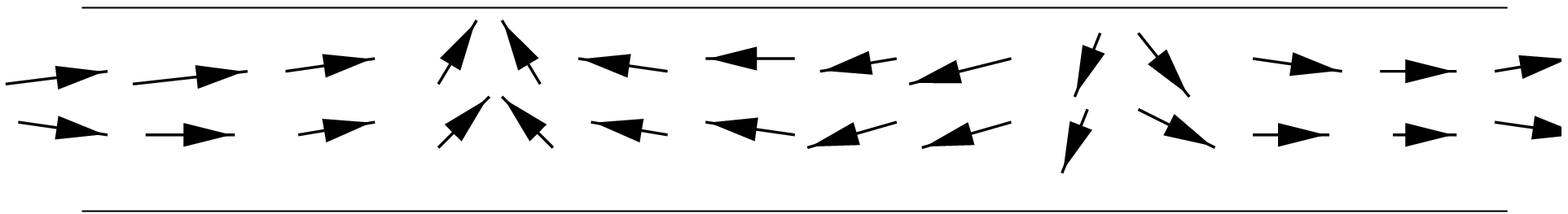}}}
\caption{Vortex field strength after maximal abelian gauge fixing.
Vortex field strength is mainly in the $\pm \s_3$ direction}
\label{avort2}
\bigskip
\bigskip
\bigskip
\centerline{\scalebox{0.5}{\includegraphics{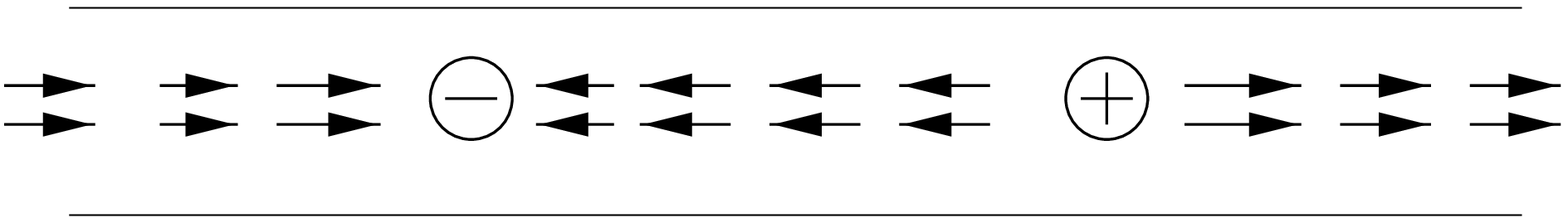}}}
\caption{Vortex field after abelian projection.}
\label{avort3}
\end{figure}

   Much of the numerical work on the center vortex theory has focused
on correlations between the location of center vortices, identified
by the center projection method, and the values of the usual gauge-invariant 
Wilson loops (cf.\ \cite{Jan98}, \cite{dFE}).  In the abelian projection 
approach, on the other hand, Wilson loops are generally computed on 
abelian projected lattices, and this fact might seem to inhibit
any direct comparison of the monopole and vortex theories.
However, it has also been suggested in ref.\ \cite{Zako} that a center
vortex would appear, upon abelian projection, in the form of a
monopole-antimonopole chain, as indicated very 
schematically in Fig.\ \ref{avort5}.
The idea is to consider, at fixed time, the vortex color-magnetic 
field in the vortex direction.
In the absence of gauge fixing, the vortex field points in arbitrary 
directions in color space, as shown in Fig.\ \ref{avort1}.  Upon fixing
to maximal abelian gauge, the vortex field tends to line up, in color 
space, mainly in the $\pm \s^3$ direction.  
But there are still going to be regions 
along the vortex tube where the field rotates from the
$+\s_3$ to the $-\s_3$ direction in color space (Fig. \ref{avort2}).  
Upon abelian projection, these regions show up as monopoles or 
antimonopoles, as illustrated in Fig.\ \ref{avort3}.  If this picture is right,
then the $\pm 2 \pi$ monopole flux is not distributed symmetrically
on the abelian-projected lattice, as one might expect in a Coulomb gas.  
Rather, it will be collimated in units of $\pm \pi$ along the vortex line.  
We have argued elsewhere \cite{GG3} 
that some sort of collimation of monopole magnetic fields into units 
of $\pm \pi$  is likely to occur even in the $D=3$ Georgi-Glashow 
model, albeit on a scale which increases exponentially with the mass of the 
W-boson.  On these large scales, the ground 
state of the Georgi-Glashow model cannot be adequately represented by 
the monopole Coulomb gas analyzed by Polyakov in ref.\ \cite{Polyakov}. The 
question we address here is whether such flux collimation also
occurs on the abelian-projected lattice of $D=4$ pure Yang-Mills theory.

   The test of flux collimation on abelian-projected lattices is in
principle quite simple.  Consider a very large abelian Wilson loop 
\beq
       W_q(C) = \Big< \exp[iq \oint dx^\m A_\m] \Big>
\label{loop}
\eeq
or abelian Polyakov line
\beq
       P_q = \Big< \exp[iq \int dt A_0] \Big>
\label{line}
\eeq
corresponding to $q$ units of the electric charge.  The expectation
values are obtained on abelian-projected lattices, extracted in maximal
abelian gauge.\footnote{Abelian-projected links in Yang-Mills theory
are diagonal matrices of the form 
$U_\mu = \mbox{diag}[\exp(iA_\mu),\exp(-iA_\mu)]$.}
If $q$ is an even number, then magnetic flux of magnitude $\pm \pi$
through the Wilson loop will not affect the loop.  Flux collimation therefore
implies that $W_q(C)$ has an asymptotic perimeter-law falloff if
$q$ is even.  Likewise, Polyakov lines $P_q$ for even $q$ are not 
affected, at long range, by collimated vortices of $\pm \pi$ magnetic flux.  
In the confined phase, the prediction is that $P_q=0$ only for odd-integer $q$.
In contrast, we expect in a monopole Coulomb gas of the sort
analyzed by Polyakov that $W_q(C)$ has an area-law falloff,
and $P_q=0$, for all $q$.

   Numerical results for abelian-projected Wilson loops and Polyakov
lines must be interpreted with some caution in the usual maximal abelian 
gauge, due to the absence of a transfer matrix in this gauge.  Since 
positivity is not guaranteed, these expectation values need not relate 
directly to the energies of physical states.  For this reason, we prefer to 
interpet the abelian observables in terms of the type of global symmetry,
or type of ``magnetic disorder'', present in the abelian-projected lattice,
without making any direct reference to the potential between abelian 
charges, or the energies of isolated charges.  Following ref.\ \cite{GG3},  
let us introduce the U(1) holonomy probability distribution on 
abelian-projected lattices  
\beq
{\cal P}_C[g] = \Big< \d \left[ g , \exp(i \oint dx^\m A_\m)\right] \Big>
\label{PC}
\eeq
for Wilson loops, and
\beq
     {\cal P}_T[g] = \Big< \d\left[ g , \exp(i \int dt A_0)\right] \Big>
\label{PT}
\eeq
for Polyakov lines, where
\beq
    \d \Bigl[ e^{i\th_1},e^{i\th_2} \Bigr] = {1 \over 2\pi}
         \sum_{n=-\infty}^{\infty} e^{in(\th_2-\th_1)}
\label{d}
\eeq
is the $\d$-function on the U(1) manifold. 
These distributions give us the probability density
that a given abelian Wilson loop around curve $C$, or an abelian 
Polyakov line of length $T$, respectively, will be found to have the 
value $g \in U(1)$ in any thermalized, abelian-projected lattice.  The 
lattice has global U(1) symmetry (or, in alternative 
terminology, the lattice has ``U(1) magnetic disorder'') if these 
distributions are flat for Polyakov lines and large Wilson loops.  In other 
words, we have U(1) symmetry iff, for any $g,g' \in U(1)$, it is true that
\beq
       {\cal P}_T[g] -  {\cal P}_T[g'g] = 0
\label{gU1}
\eeq
for Polyakov lines, and
\beq
       {\cal P}_C[g] -  {\cal P}_C[g'g] \sim \exp[-\s \mbox{Area}(C)]
\eeq
holds asymptotically for large Wilson loops.  In the case that these relations 
are not true in general, but the restricted forms
\beq
       {\cal P}_T[g] -  {\cal P}_T[zg] = 0
\label{gZ2}
\eeq
and
\beq
       {\cal P}_C[g] -  {\cal P}_C[zg] \sim \exp[-\s \mbox{Area}(C)]
\eeq
hold for any $g\in U(1)$ and $z = \pm 1 \in Z_2$,  then we will say that
the lattice has only $Z_2$ global symmetry (or $Z_2$ magnetic 
disorder).\footnote{Generalizing to $g \in SU(N)$ and $z \in Z_N$, for
gauge-invariant loops on the unprojected lattice,
eq.\ \rf{gZ2} follows from the well-known global $Z_N$ symmetry of the confined
phase.} 

  Inserting \rf{d} into \rf{PC} and \rf{PT}, we have 
\bea
     {\cal P}_C[e^{i\th}] &=& {1\over 2\pi}\left(1 + 
2 \sum_{q>0} W_q(C) \cos(q\th) \right)  
\non \\
     {\cal P}_T[e^{i\th}] &=& {1\over 2\pi}\left(1 + 
2 \sum_{q>0} P_q \cos(q\th) \right)
\label{expand}
\eea
From this, we can immediately see that there is U(1) magnetic disorder
iff $W_q(C)$ has an asymptotic area law falloff, and $P_q=0$, for all
integer $q \ne 0$.  On the other hand, if these conditions hold only for
$q=$odd integer, then we have $Z_2$ magnetic disorder.  It is an
unambiguous prediction of the vortex theory that the lattice has
only $Z_2$ magnetic disorder, even after abelian projection.

   The magnetic disorder induced by a monopole Coulomb gas is expected
to be rather different from the disorder induced by vortices.  A Coulombic 
magnetic field distribution will 
in general affect loops of any $q$, with $q>1$ loops responding even more 
strongly than $q=1$ to any variation of magnetic flux through the loop.  
The usual statement is that all integer abelian charges are confined.
This statement is confirmed explicitly in $D=3$ dimensions, where 
it is found in a
semiclassical calculation that the monopole Coulomb gas derived from
$QED_3$ confines all charges, with string tensions $\s_q$ directly 
proportional to the charge \cite{GG3}
\beq
       \s_q = q \s_1
\label{mcg}
\eeq
This relation is also consistent with recent numerical simulations 
\cite{Manfried}.  

   In D=4 dimensions, an analytic treatment of monopole currents interacting
via a two-point long-range Coulomb propagator, plus possible contact terms, 
is rather difficult.  
Nevertheless, $QED_4$ (particularly in the Villain formulation) can be 
viewed as a theory of monopole loops and photons, and in the 
confined (=strong-coupling)
phase there is a string tension for all charges $q$, and all $P_q=0$,
as can be readily verified from the strong-coupling expansion.  Confinement
of all charges $q$ is also found in a simple model of the monopole Coulomb 
gas, due to Hart and Teper \cite{HT}.  Finally, all multiples $q$ of
electric charge are confined in the dual abelian Higgs model (a theory of
dual superconductivity), and this model is
known \cite{Maxim} to be equivalent, in certain limits, to an
effective monopole action with long-range two-point Coulombic interactions 
between monopole currents.  

   In the case of $D=4$ abelian-projected Yang-Mills theory, we do not
really know if the distribution of monopole loops identified on the projected
lattice is typical of a monopole Coulomb gas.  What \emph{can} be tested,
however,  is whether or not the field associated with these monopoles
is Coulombic (as opposed, e.g., to collimated). This is done by comparing
observables measured on abelian-projected lattices with those obtained
numerically via a ``monopole dominance'' (MD)
approximation, first introduced in ref.\ \cite{MD}. The MD approximation 
involves two steps.  First, the location of 
monopole currents on an abelian-projected lattice is identified using the
standard DeGrand-Toussaint criterion \cite{dGT}.  Secondly, a lattice 
configuration is reconstructed by assuming that each link is affected by 
the monopole currents via a lattice Coulomb propagator.  Thus, the lattice
Monte Carlo and abelian projection supply a certain distribution of monopoles,
and we can study the consequences of 
assigning a Coulombic field distribution to the monopole charges.

   We should pause here, to explain how the non-confinement of 
color charges in the adjoint representation is accounted for in the
monopole gas or dual-superconductor pictures, in which all abelian charges
are confined. The screening of adjoint (or, in general, $j=$integer) 
representations in Yang-Mills theory is a consequence of having only
$Z_2$ global symmetry on a finite lattice in the confined phase.
In a monopole Coulomb gas, on the other hand, the confinement of all
abelian charges would imply a U(1) global symmetry of the projected
lattice at finite temperature.  Nevertheless, the $Z_2$ symmetry of
the full lattice and U(1) symmetry of the projected lattice are not
\emph{necessarily} inconsistent.
Let $P^{YM}_j$ represent the Polyakov line in SU(2) gauge theory in group
representation $j$.  Generalizing eq.\ \rf{PT} to $g \in SU(2)$, we have
\beq
     {\cal P}_T[g] = \sum_{j=0,\oh,1...} P^{YM}_j \chi_j[g]
\eeq         
and the fact that eq.\ \rf{gZ2}, rather than eq.\ \rf{gU1}, is satisfied
follows from the identity
\beq
      \chi_j(zg) = \chi_j(g) ~~~~\mbox{for~} j = \mbox{integer}
\eeq
and from the fact that 
\beq
      P^{YM}_j  \left\{ \begin{array}{rl}
              = 0 & j = \mbox{half-integer} \cr
            \ne 0 & j = \mbox{integer} > 0 \end{array} \right.
\eeq
in the confined phase, due to confinement of charges in half-integer, 
and color-screening of charges in integer, SU(2) representations.

   Now, on an abelian projected lattice, the expectation value of a
Yang-Mills Polyakov line in representation $j$ becomes
\beq
      P^{YM}_j \ra \sum_{m=-j}^j P_{2m}
\eeq
where $P_q$ is the abelian Polyakov line defined in eq.\ \rf{line}.
If the abelian-projected lattice is U(1) symmetric, then $P_{2m}=0$
for all $m \ne 0$, while $P_0=1$.  This means that after abelian-projection
we have
\beq  
      P^{YM}_j = \left\{ \begin{array}{rl}
               0 & j = \mbox{half-integer} \cr
               1 & j = \mbox{integer} \end{array} \right.
\eeq
and in this way the non-zero values of $j=$integer Polyakov lines
are accounted for, even assuming that the projected lattice  has a global
U(1)-invariance.  Similarly, adjoint Wilson loops will not have an
area law falloff on the projected lattice, as expected from color-screening.
This explanation of the adjoint perimeter law in the abelian projection
theory has other difficulties, associated with Casimir scaling
of the string tension at intermediate distances (cf. ref.\ \cite{Lat96}), 
but at least the \emph{asympotic} behavior of Wilson loops in various 
representations is consistent with U(1) symmetry on the abelian projected 
lattice.  

   It is just this assumed U(1) symmetry of the abelian-projected lattice, and
the monopole Coulomb gas picture which is associated with it, that we
question here.  The U(1) vs.\ $Z_2$ global symmetry issue
can be settled by calculating abelian Wilson loops and/or Polyakov lines
for abelian charges $q>1$.  The validity of the Coulomb gas picture can
also be probed by calculating $W_q(C)$ and $P_q$ with and without the monopole
dominance (MD) approximation, and comparing the two sets of quantities.  

   The first calculation of $q>1$ Wilson loops, in the MD approximation,
was reported recently by Hart and Teper \cite{HT}; their calculation
confirms the Coulomb gas relation $\s_q \propto q$ found previously
for compact $QED_3$.  In ref.\ \cite{HT}, this result is interpreted
as favoring the monopole Coulomb gas picture over the vortex theory.
From our previous remarks, it may already be clear why we do not
accept this interpretation.  At issue is whether monopole magnetic fields
spread out as implied by the Coulomb propagator, or whether they are
collimated in units of $\pm \pi$.  This issue cannot be resolved by
the MD approximation, which imposes a Coulombic field distribution from the
beginning.  The MD approximation does, however, tell us that if the
monopoles have a Coulombic field distribution, then the $q=2$ Wilson loop
has an area law falloff, at least up to the maximum charge separation
studied in ref.\ \cite{HT}.  The crucial question is whether the $q=2$ 
loops computed directly on abelian projected lattices also have an asymptotic 
area law falloff, or instead go over to perimeter-law behavior (usually
called ``string-breaking'') as predicted by the vortex theory.
   
   Here it is important to have some rough idea of where the $q=2$
string is expected to break, according to the vortex picture,
otherwise a null result can never be decisive.  A $q=2$ Wilson loop 
will go over to perimeter behavior when the size of the loop is 
comparable to the thickness of the vortex.  It seems reasonable to
assume that the thickness of a vortex on the abelian-projected lattice 
is comparable to the thickness of a center vortex on the unprojected lattice.
From Fig.\ 1 of ref.\ \cite{Jan98}, this thickness appears to be roughly 
one fermi, which is also about the distance where an adjoint representation
string should break in $D=4$, according to an estimate due to Michael 
\cite{Michael}.  The finite thickness of the vortex is an important 
feature of the vortex theory,
as it allows us to account for the approximate Casimir scaling of string 
tensions at intermediate distances (cf. refs.\ \cite{Cas1,Cas2}).  But it 
also means that at, e.g., $\b=2.5$, we should look for string breaking 
at around $R=12$ lattice spacings.  Noise reduction techniques, such as the
``thick-link'' approach, then become essential.

   The validity of the thick-link approach, however, is tied to the
existence of a transfer matrix.  Since the method uses $R\times T$ loops
with $R\gg T$, one has to show that the potential extracted is mainly 
sensitive to the large separation $R$, rather than the smaller separation
$T$, and here positivity plays a crucial rule.  Since there is no transfer
matrix in maximal abelian gauge, the validity of the thick link approach
is questionable (and the issue of positivity is much more than a quibble, as
we will see below).  Moreover, even when a transfer matrix exists, 
string-breaking is not easy to observe by this method, and requires more 
than just the calculation of rectangular loops.  
The breaking of the adjoint-representation string has not been seen
using rectangular loops alone, and only quite recently has this
breaking been observed, in 2+1 dimensions, by taking account of
mixings between string and gluelump operators \cite{Stephenson,Wittig}.  
The analogous calculation, for operators defined in maximal abelian
gauge, would presumably involve mixings between
the $q=2$ string and ``charge-lumps''; the latter being bound states
of the static abelian charge and the off-diagonal 
(double abelian-charged) gluons.  

   There are, in fact, existing calculations of the $q=2$ potential, 
by Poulis \cite{Poulis} and by Bali et al.\ \cite{Bali}, using
the thick-link method.  String breaking was not observed, but neither
did these calculations make use of operator-mixing techniques, which 
seem to be necessary for this purpose.  In any case, in view of the absence
of a transfer matrix in maximal abelian gauge, we do not regard these
calculations as decisive.

   Given our reservations concerning the thick-link approach,
we will opt in this article for a far simpler probe of 
global symmetry/magnetic disorder on the projected lattice, 
namely, the double-charged abelian Polyakov lines $P_2$.  Any 
abelian magnetic vortex
can be regarded, away from the region of non-vanishing vortex field
strength, as a discontinuous gauge transformation, and it is this
discontinuity which affects Wilson loops and Polyakov lines far
from the region of finite vortex field strength.
If the vortex flux is $\pm \pi$, the discontinuity will not affect
even-integer $q$-charged Polyakov lines, and these should have a
finite expectation value. The abelian-projected lattice then has only 
$Z_2$ global symmetry in the confined phase.  In contrast, a monopole 
Coulomb gas is expected to confine all $q$ charges, as in compact $QED_3$, 
and the $Z_2$ subgroup should play no special role. 
In that case, $P_q=0$ for all $q$.
Thus, if we find that the $q=2$ Polyakov line vanishes, this is evidence 
against vortex structure and flux collimation, and in favor of the
monopole picture.  Conversely, if $q=2$ Polyakov lines do not vanish, 
the opposite conclusion applies, and the vortex theory is favored.

\setcounter{equation}{0}
\section{Polyakov lines}

   After fixing to maximal abelian gauge in SU(2) lattice gauge theory, 
abelian link variables
\beq
        U^A_\m(x) = \mbox{diag}[e^{i\th_\m(x)},e^{-i\th_\m(x)}]
\label{links}
\eeq
are extracted by setting the off-diagonal elements of link variables
$U_{\mu}$ to zero, and rescaling to restore unitarity.  A $q$-charge Polyakov
line $P_q(\vx)$ is defined as
\beq
     P_q(\vx) = \prod_{n=1}^{N_T} \exp[iq\th_4(\vx+n\hat{4})]
\label{pqx}
\eeq
where $N_T$ ($N_S$) is the number of lattice spacings in the 
time (space) directions.
We can consider both the expectation value of the lattice average
\beq
     P_q = \left< {1\over N_S^3} \sum_{\vx} P_q(\vx) \right>
\label{vev}
\eeq
and the expectation value of the absolute value of the lattice
average
\beq
   P^{abs}_q = \left< {1\over N_S^3}\left| \sum_{\vx} P_q(\vx)\right| \right>
\label{abs}
\eeq
Polyakov lines can vanish for $q=1$, even in the deconfined phase, just
by averaging over $Z_2$-degenerate vacua, which motivates the absolute
value prescription.  In the confined phase, one then has 
$P^{abs}_1 \sim N_S^{-3/2}$.
We will find that this prescription is unnecessary 
for $q=2$, and we will compute these Polyakov lines 
without taking the absolute value of the lattice average.

   For purposes of comparison, and as a probe of the monopole Coulomb
gas picture, we also compute ``monopole'' Polyakov 
lines $P_{Mq}$ following an MD approach used by Suzuki et al.\ in ref.\ 
\cite{MD1}. Their procedure is to decompose the abelian plaquette 
variable ($\pa_\m$ denotes the forward lattice difference)
\beq
       f_{\m \n}(x) = \pa_\m \th_\n(x) - \pa_\n \th_\m(x)
\label{MD1}
\eeq
into two terms
\beq
   f_{\m \n}(x) = \overline{f}_{\m \n}(x) + 2\pi n_{\m \n}(x)
\label{MD2}
\eeq  
where $n_{\m \n}$ is an integer-valued Dirac-string variable, and 
$-\pi < \overline{f}_{\m\n} \le \pi$.  One can then invert \rf{MD1}
to solve for $\th_4$ in terms of the ``photon'' field-strength
$\overline{f}_{\m\n}$, the Dirac-string variables $n_{\m\n}$, and an
irrelevant U(1) gauge-dependent term.
If we assume that the photon and Dirac-string variables are completely
uncorrelated, then the Dirac-string contribution is given by
\beq
     \th^M_4(x) = - \sum_{x'} D(x,x') \pa'_\n n_{\n 4}(x')
\label{MD3}
\eeq
Here $D(x,x')$ is the lattice Coulomb propagator, and
the partial derivative denotes a backward difference.  The
monopole dominance approximation is to replace $\th_4$ by
$\th^M_4$ in eq.\ \rf{pqx}, the idea being that this procedure isolates
the contribution of the monopole fields to the Polyakov lines. The
correlations between the photon, monopole, and abelian lattice fields
will be discussed in more detail in section 3, and in an Appendix.

\subsection{\boldmath $Z_2$ Magnetic Disorder}

\begin{figure}[h!]
\centerline{\scalebox{0.75}{\includegraphics{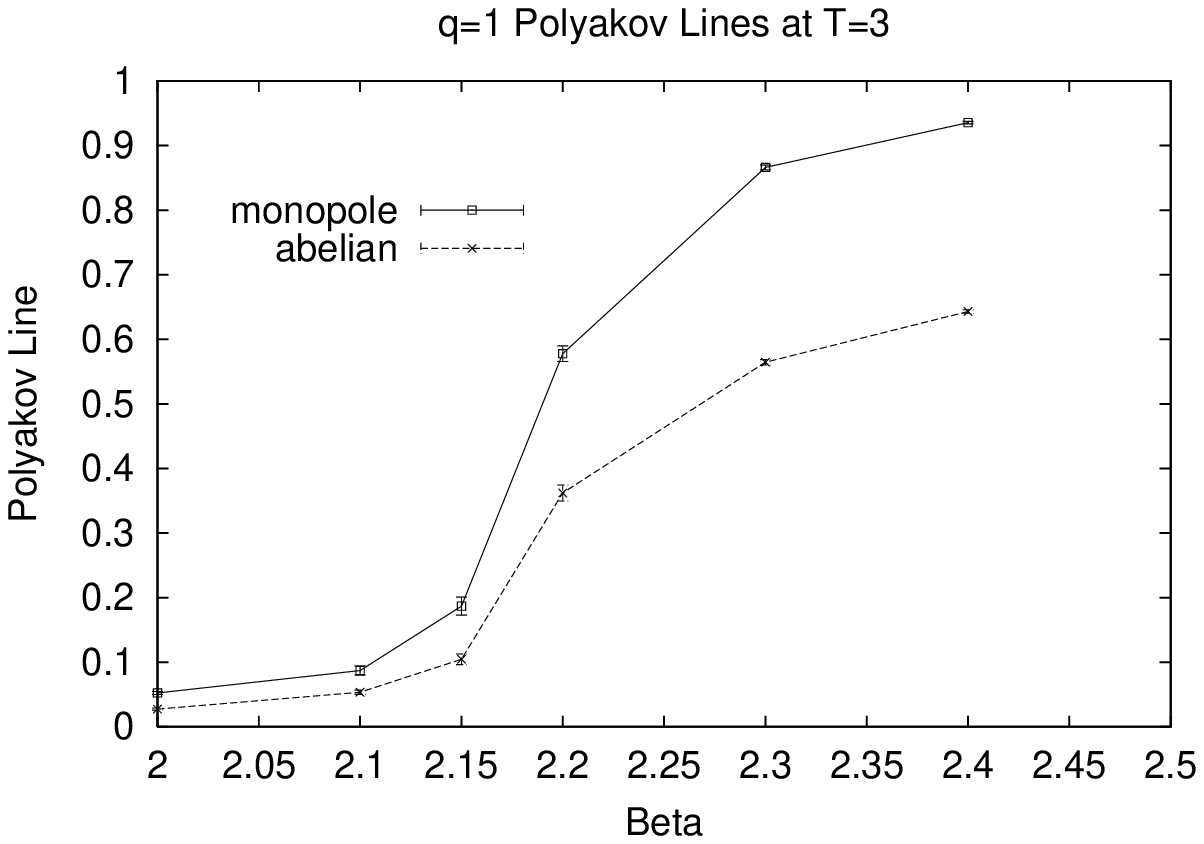}}}
\caption{Single charge ($q=1$) Polyakov lines at time extention T=3, 
on the abelian projected lattice and in the monopole
dominance (MD) approximation.}
\label{P1T3}
\end{figure}

   In Fig.\ \ref{P1T3} we display $P^{abs}_1$ and $P^{abs}_{M1}$ for the
$q=1$ lines, on a $12^3 \times 3$ lattice.  There are no surprises here; 
we see that for $N_T=3$ there is a deconfinement
transition around $\b=2.15$.

\begin{figure}[h!]
\centerline{\scalebox{0.75}{\includegraphics{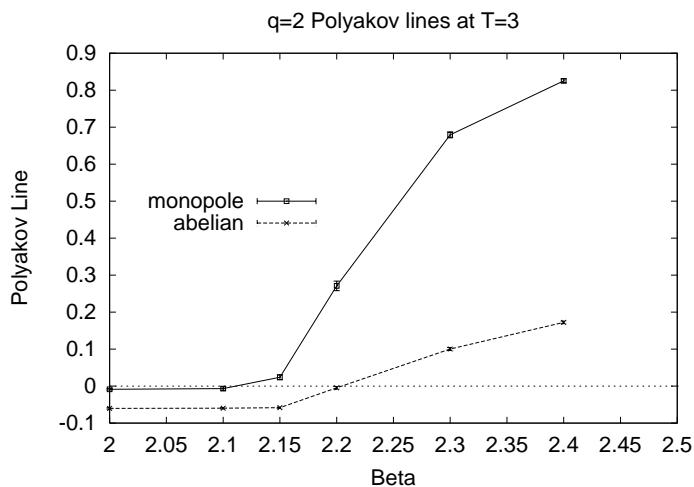}}}
\caption{Doubly-charged ($q=2$) Polyakov lines at time extension T=3, 
on the abelian projected lattice and in the monopole
dominance (MD) approximation.}
\label{P2T3}
\end{figure}

\begin{figure}[h!]
\centerline{\scalebox{0.75}{\includegraphics{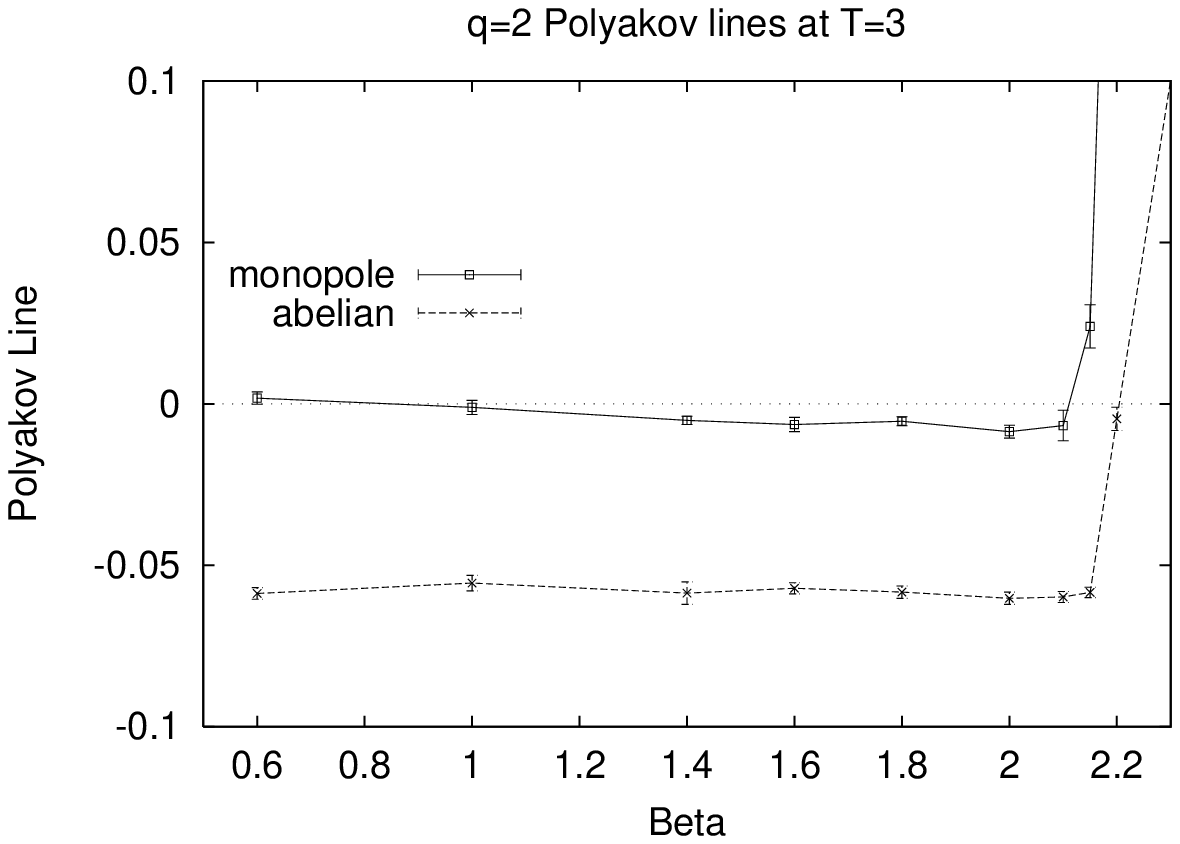}}}
\caption{Magnified view of $q=2$ abelian and monopole-dominance 
Polyakov lines at time extension $T=3$, in the confined phase.}
\label{P2T3a}
\end{figure}

   The situation changes dramatically when we consider $q=2$ Polyakov
lines.  Fig.\ \ref{P2T3} is a plot of the values of $P_2$ and $P_{M2}$, without
any absolute value prescription, on the $12^3 \times 3$ lattice.  To make 
the point clear, we focus on the data in the confined phase, in Fig.\ 
\ref{P2T3a}. It can be seen that $P_2$ {\it is non-vanishing and negative 
in the confined phase}; the data is clearly not consistent with a vanishing 
expectation value.  In the MD approximation, $P_{M2}$
may also be slightly negative, but its value is at least
an order of magnitude smaller than $P_2$.  This seems to be a very strong 
breakdown of monopole dominance, in the form proposed in ref.\ \cite{MD1}.

\begin{figure}[h!]
\centerline{\scalebox{0.75}{\includegraphics{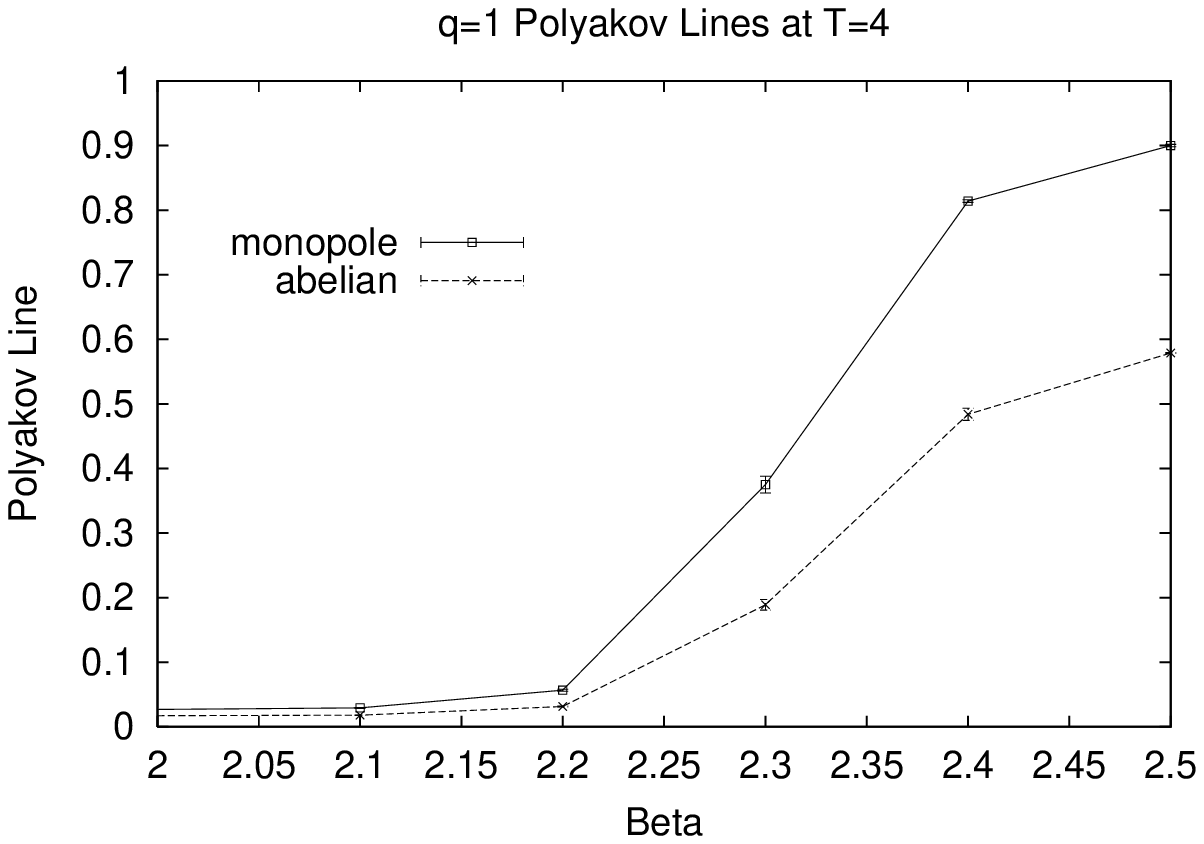}}}
\caption{$q=1$ Polyakov lines (as in Fig.\ \ref{P1T3}), 
for time extension T=4 lattice spacings.}
\label{P1T4}
\end{figure}

\begin{figure}[h!]
\centerline{\scalebox{0.75}{\includegraphics{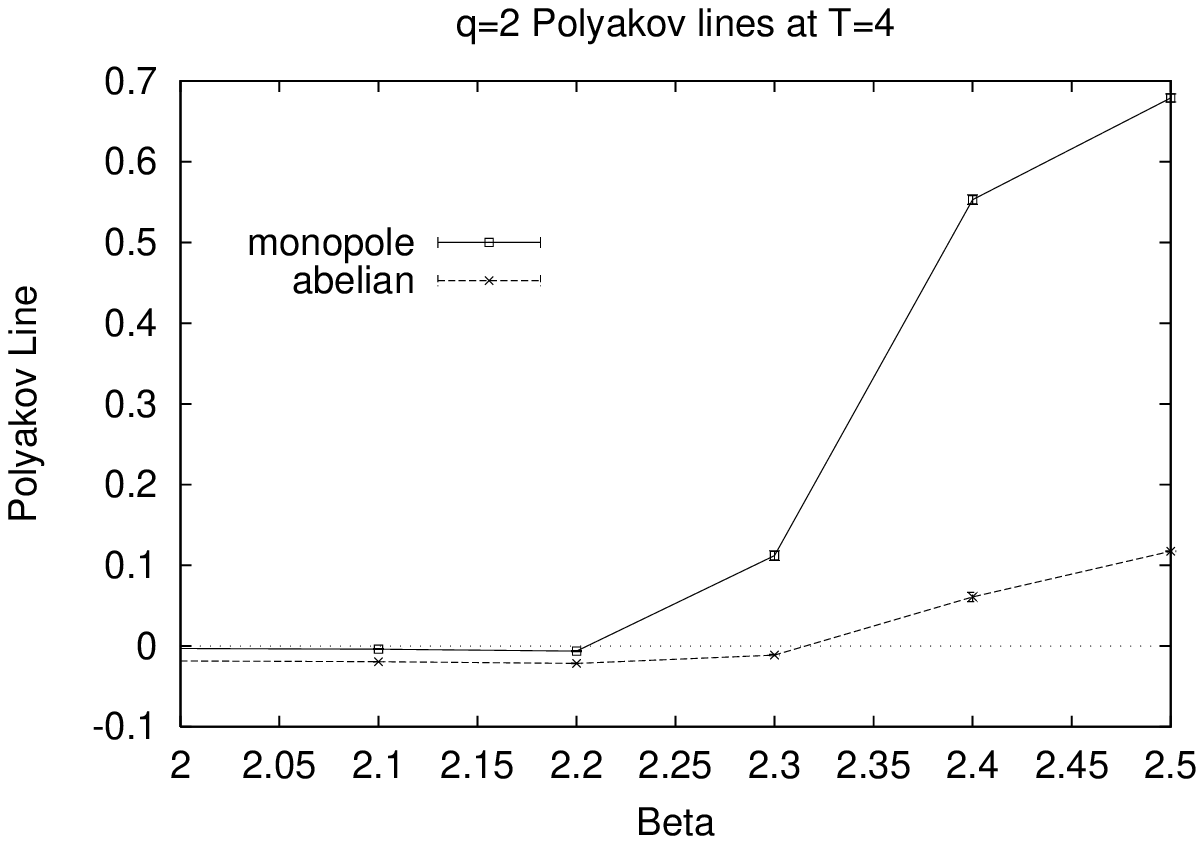}}}
\caption{$q=2$ Polyakov lines (as in Fig.\ \ref{P2T3}), for time extension T=4
lattice spacings.}
\label{P2T4}
\end{figure}

\begin{figure}[h!]
\centerline{\scalebox{0.75}{\includegraphics{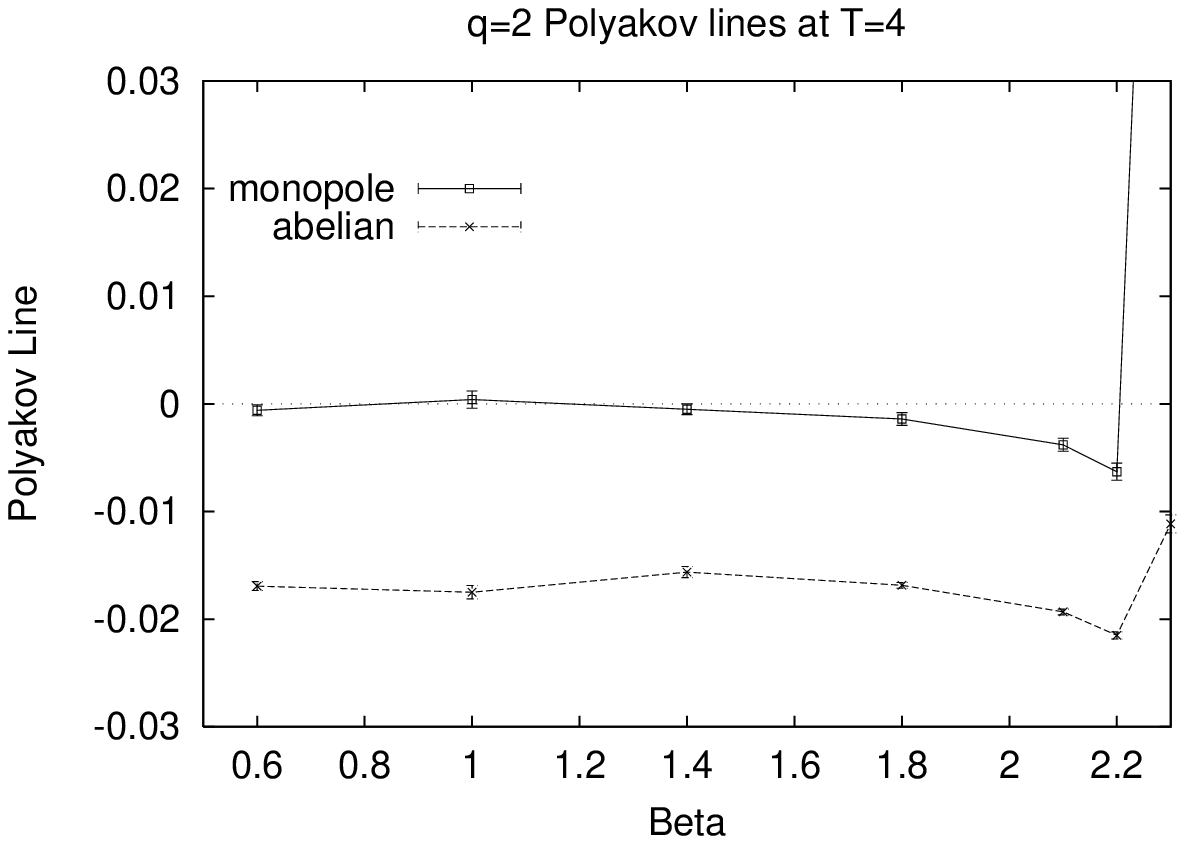}}}
\caption{$q=2$ Polyakov lines in the confined phase (as in Fig.\ \ref{P2T3a}), 
for time extension T=4 lattice spacings.}
\label{P2T4a}
\end{figure}

  In Figures \ref{P1T4}-\ref{P2T4a} we plot the corresponding data
found on a $16^3\times 4$ lattice.  There is a deconfinement transition
close to $\b=2.3$, and again there is a clear disagreement between
$P_2$ and $P_{M2}$, with the former having a substantial non-vanishing
expectation value throughout the confined phase.

  The numerical evidence, for both $T=3$ and $T=4$, clearly favors 
having $Z_2$, rather than U(1), global symmetry/magnetic disorder on the 
abelian projected lattice.

\subsection{Spacelike Maximal Abelian Gauge}

\begin{figure}[h!]
\centerline{\scalebox{0.75}{\includegraphics{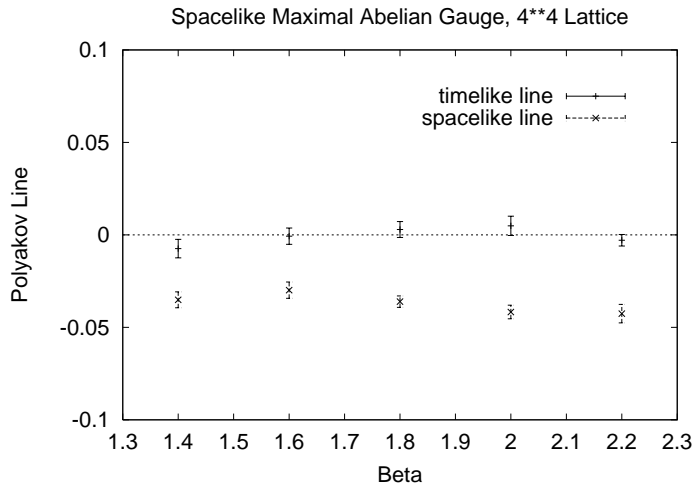}}}
\caption{Breaking of $90^\circ$ rotation symmetry in spacelike
maximal abelian gauge, as seen from comparing timelike Polyakov lines 
(lattice 4 direction) with spacelike Polyakov lines 
(lattice 1,2, or 3 directions) on a $4^4$ lattice.}
\label{polyc}
\end{figure}

   It is also significant that $P_2$ is negative.  This
implies a lack of reflection positivity in the Lagrangian obtained after 
maximal abelian gauge fixing, and must be tied to the fact that maximal
abelian gauge is not a physical gauge.  This diagnosis also suggests
a possible cure: Instead of fixing to the standard maximal abelian gauge,
which maximizes
\beq
 R = \sum_x \sum_{\m=1}^4 \mbox{Tr} [\s_3 U_\m(x) \s_3 U^\dg_\m(x)]
\eeq
we could try to use a ``spacelike'' maximal abelian gauge \cite{Polikarp}, 
maximizing the quantity
\beq
 R = \sum_x \sum_{k=1}^3 \mbox{Tr} [\s_3 U_k(x) \s_3 U^\dg_k(x)]
\eeq
which involves only links in spatial directions.  This is a physical gauge.
What happens in this case is that one disease, the loss of reflection
positivity, it replaced by another, namely, the breaking of $90^\circ$
rotation symmetry.  This is illustrated in Fig.\ \ref{polyc}, where we plot 
spacelike and timelike Polyakov lines on a $4^4$ lattice, in the 
spacelike maximal abelian gauge defined above.  We find that
the values for double-charged Polyakov lines running in the time direction
are much reduced in the spacelike gauge, and in fact the results shown
appear consistent with zero.  \emph{Spacelike} $q=2$ Polyakov lines, however, 
which run along the $1,2,$ or $3$ lattice directions, remain negative, and 
in fact are larger in magnitude than Polyakov lines of the same length, 
and the same coupling, computed in the usual maximal abelian gauge.  One 
therefore finds on a hypercubic lattice that $90^\circ$ rotation symmetry 
is broken.

   The spacelike Polyakov line operator creates a line of electric flux
through the periodic lattice.  The non-vanishing overlap of this state
with the vacuum has, in the spacelike gauge, a direct physical interpretation:
Since the $q=2$ electric flux line cannot, for topological reasons, shrink 
to zero,  a finite overlap with the vacuum means that the $q=2$ flux tube
breaks.  This is presumably due to screening by double-charged 
(off-diagonal) gluon fields.  The implication is that in a physical gauge,
where Wilson loops can be translated into statements about potential 
energies, $q=$ even abelian charges are not confined.

   The fact that $q=$ even charges are unconfined, together with the
positivity property in spacelike maximal abelian gauge, leads to the
conclusion that timelike $q=2$ Polyakov lines $P_2$ are positive and
non-zero, although the data points for timelike $P_2$ shown in 
Fig.\ \ref{polyc}, which appear to be consistent with zero,
do not yet support such a conclusion.  It must be that the value of
$P_2$ in this gauge is simply very small, and much better statistics are
required to distinguish that value from zero.  To get some idea of the
difficulty involved, let us suppose that the magnitude of the timelike
abelian line $P_2$ on the projected lattice is comparable to the magnitude 
of the gauge-invariant Polyakov line $P_{adj}$, in the adjoint representation 
of SU(2), on the full, unprojected lattice.  To leading order in the
strong-coupling expansion, $P_{adj}$ on a lattice with extension $T$ in
the time direction is given by
\beq
      P_{adj} = 4 \left( {\b \over 4} \right)^{4T}
\eeq
This equals, e.g., $0.00156$ for $T=2$ at $\b=1.5$; quite a small signal
considering that $T$ is only two lattice units.  The obvious remedy is 
to increase $\b$, but then one runs into a deconfinement transition at 
$\b=1.8$.  We can move the transition to larger $\b$ by increasing $T$, 
but of course increasing $T$ again causes the signal to go down.

   The best chance to extract a signal from the noise is to choose a
value of $\b$ which is fairly close to the deconfinement transition
(but still in the confined phase), and to generate very many configurations.
Here are the results obtained at $T=2$ lattice spacings and 
$\b=1.7$, coming from 5000
configurations separated by 100 sweeps on a $2 \times 8^3$ lattice:
\bea
      P_{adj} &=&  0.00447(23)
\non \\
      P_2     &=&  0.00241(52)     
\eea
The result for the adjoint line is consistent with the 
strong-coupling prediction of $P_{adj}=0.00426$.  The abelian line $P_2$ is
non-zero, positive, and comparable in magnitude to $P_{adj}$, although 
it must be admitted that the errorbar is uncomfortably large.  The
corresponding values for $T=3$ and $\b=2.11$, obtained from 5000
$3\times 8^3$ lattices, are
\bea
      P_{adj} &=&  0.00338(28)
\non \\
      P_2     &=&  0.00124(42)     
\eea
Again $P_2$ is non-zero, although the errorbar is still 
too large for comfort.  

   Clearly the evaluation of the timelike $P_2$ line in spacelike
maximal abelian gauge is cpu-intensive, and our results 
for this quantity must be regarded as preliminary.  Nevertheless,
these preliminary results are consistent with the conclusion previously 
inferred from the spacelike lines: In a physical gauge, the $q=2$
charge is screened, rather than confined, and we have $Z_2$, rather
than U(1), magnetic disorder on the abelian-projected lattice.

\setcounter{equation}{0}
\section{The ``Photon'' Contribution}

   Suppose we write the link angles $\th_\m(x)$ of the abelian link
variables as a sum of the link angles $\th^M_\m(x)$ in the MD 
approximation, plus a so-called ``photon'' contribution 
$\th^{ph}_\m(x)$, i.e.
\beq
       \th^{ph}_\m(x) \equiv \th_\m(x) - \th_\m^M(x)
\eeq
It was found in refs.\ \cite{MD,MD1,HT1} that the photon field has no
confinement properties at all; the Polyakov line constructed from
links $U_\mu=\exp[i\th^{ph}_\m]$ is finite, and corresponding
Wilson loops have no string tension.  Since $\th_\m^M$ would appear
to carry all the confining properties, a natural conclusion is that
the abelian lattice is indeed a monopole Coulomb gas.

   To see where this reasoning may go astray, suppose we perversely
\emph{add}, rather than subtract, the MD angles to the abelian
angles, i.e.
\bea
       \th'_\m(x) &=& \th_\m(x) + \th_\m^M(x)
\non \\
                  &=& \th^{ph}_\m(x) + 2\th_\m^M(x)
\label{add}
\eea
in effect doubling the strength of the monopole Coulomb field.
It is natural to expect a corresponding increase of the string
tension, and of course $P_1=0$ should remain true.  Surprisingly, 
this is not what happens; doubling the strength of the monopole field in fact
removes confinement.\footnote{We have already noted
that in the absence of a transfer matrix, the term ``confinement
of abelian charge'' must be used with caution.  In this section,
the phrase ``confinement of charge q'' is just taken to mean
``$P_q=0$''.} Some results for $P_1$ are shown in Table \ref{table1}.
Here we have computed the vev of $P_1(x)$ without taking the absolute
value of the lattice sum (i.e.\ we use eq.\ \rf{vev} rather than 
eq.\ \rf{abs}), and we find that $P_1$ is finite and negative in the additive
configurations.  The additive configuration $\th'$ is far from pure-gauge,
and the vev of $P_1$ is correspondingly small. Nevertheless, $P_1$ is
non-zero, so adding the monopole field in this case actually 
\emph{removes} confinement.  Clearly, the interplay between the MD 
and ``photon'' contributions is a little more subtle than previously supposed.

\begin{table}[h!]
\centerline{
\begin{tabular}{|l|r|r|} \hline\hline
  T       & $\b$  &  $P_1$ line \\ \hline
  3       &  1.8  &  -0.0299 (20) \\ 
  3       &  2.1  &  -0.0405 (10) \\
  4       &  2.1  &  -0.0134 (10) \\ \hline
\end{tabular} }
\caption{Single-charged abelian Polyakov lines, computed
in the additive $\th' = \th + \th^M$ configurations, in which the
strength of the MD contribution is effectively doubled.} 
\label{table1}
\end{table}

\begin{figure}[h!]
\centerline{\scalebox{0.50}{\includegraphics{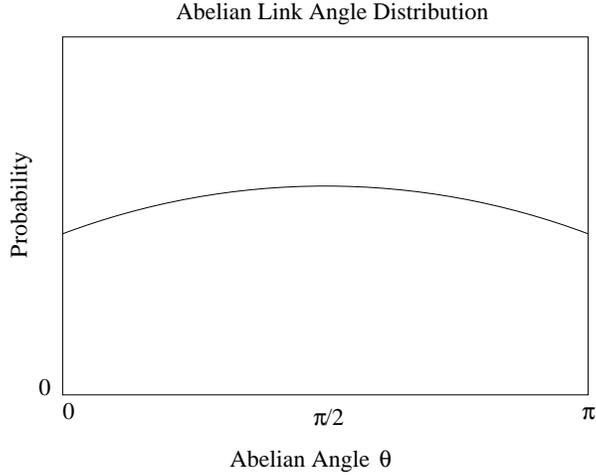}}}
\caption{Conjectured $\P(\th)$ distribution of the Polyakov
phase angle, based on $Z_2$ global symmetry and $P_2<0$.}
\label{Ptheta}
\end{figure}

\begin{figure}[h!]
\centerline{\scalebox{0.50}{\includegraphics{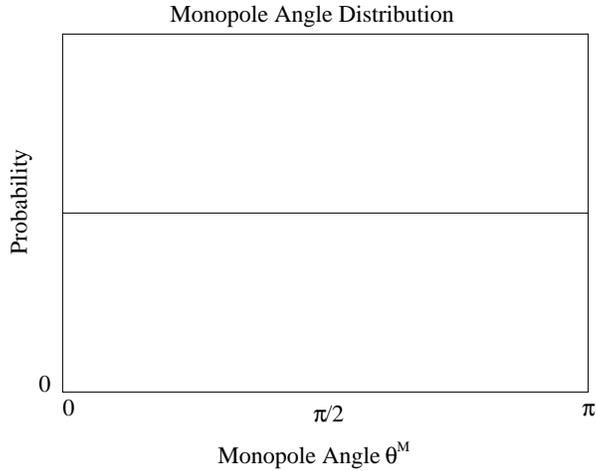}}}
\caption{$\P(\th^M)$ in the MD approximation, assuming perfect
global U(1) symmetry.}
\label{Pmpole}
\end{figure}

   To understand what is going on, we return to the concept of 
the holonomy probability distribution
\beq
     \P(\th) = \P_T[e^{i\th}] = 
\Big< \d\left[ e^{i\th} , \prod_{n=1}^{N_T} \exp[i\th_4(\vx+n\hat{4})]
        \right] \Big>
\eeq
$\P_T[e^{i\th}]$ is the probability density for the U(1) group elements
on the group manifold.  However, since the group measure on the U(1)
manifold is trivial (i.e.\ $d\th$), it is not hard to see that
$\P(\th) d\th$ is interpreted as the probability that
the phase of an abelian Polyakov line lies in the interval $[\th,\th+d\th]$.
In a similar way, replacing $\th_4$ by
$\th^M_4$ or $\th^{ph}_4$ on the rhs of the above equation, we can define 
the probability distributions $\P(\th^M)$ and $\P(\th^{ph})$, respectively, 
for the phases of monopole and photon Polyakov lines.  All of these
distributions have $2\pi$-periodicity, and are invariant under $\th \ra -\th$,
reflections, so we need only consider their behavior in the interval $[0,\pi]$.
Without making any further calculations, it is already possible to deduce
something about the shape of $\P(\th)$:
\begin{itemize}
\item Since all $P_q$ are small, $\P(\th)$ is fairly flat.
\item Assuming $Z_2$ symmetry, $\P(\th)$ is symmetric, in the
interval $[0,\pi]$, around $\th={\pi \over 2}$.
\item Since $P_2$ is negative, $\P(\th)$ should be larger in the
neighborhood of $\th={\pi \over 2}$ than in the neighborhood
of $\th=0$ or $\th=\pi$.
\end{itemize}
From these considerations, we deduce that $\P(\th)$ looks something
like Fig. \ref{Ptheta}.  
Similarly, since $P_{Mq} \approx 0$ in the MD approximation,
we conclude that there is very nearly U(1) symmetry in this approximation,
and $\P(\th^M)$ is almost flat, as in Fig. \ref{Pmpole}.

\begin{figure}[h!]
\centerline{\scalebox{0.50}{\includegraphics{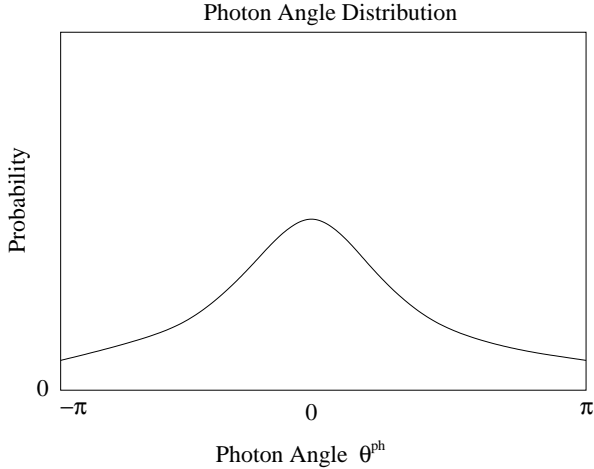}}}
\caption{The ``photon'' angle distribution $\P(\th^{ph})$, resulting
from a correlation between $\th$ and $\th^M$.}
\label{Pphoton}
\end{figure}
 
  Now if the link angles $\th_\m(x)$ and $\th^M_\m(x)$ 
are correlated to some extent,
then the difference $\th^{ph}(x)$ between these variables is not random,
but has some non-uniform probability distribution 
as illustrated in Fig. \ref{Pphoton}.  Since the Fourier 
cosine components of $\P(\th^{ph})$ are typically non-zero, it follows
that the photon field, by itself,
has no confinement property.  The crucial point is that by subtracting
$\th^M$, the $Z_2$ symmetry of $\th$ is broken due to the correlation
between $\th$ and $\th^M$.
It is interesting to note that even if
$\P(\th^M)$ were neither U(1) nor $Z_2$ symmetric, i.e.\ if we imagine
that the $\th$-configurations confine but the MD contributions do not,
a correlation between the $\th$ and $\th^M$ would still be sufficient to break
the $Z_2$ symmetry of the difference configuration $\th^{ph}$.  As a
result, subtracting the non-confining $\th^M$ from the confining $\th$ would 
still remove confinement.    

   In the center vortex picture, vortex fields supply the confining disorder,
but of course this does not at all exclude a correlation of the MD variables
$\th^M$ with the $\th$ variables.  According to the arguments in the 
Introduction,  monopoles lie along vortices as shown in
Fig.\ 1 (further evidence is given in the next section), and this 
correspondence will certainly introduce 
some degree of correlation between magnetic flux on the
abelian-projected lattice, and magnetic flux in the MD approximation.
Roughly speaking, one can say that the confining flux has the same
magnitude on the abelian and MD lattices, only it is distributed
differently (collimated vs.\ Coulombic).  However, according to
the center vortex picture, there must \emph{also} be some correlation
between $\th^{ph}_\m(x)$ and $\th^M_\m(x)$; this is necessary to
convert the long-range monopole Coulomb field into a vortex field,
and to break the U(1) symmetry of the MD lattice down to
the $Z_2$ symmetry of the vortex vacuum.  

   In numerical simulations performed at $\b=2.1$ and $T=3$ on
a $3\times 12^3$ lattice, we do, in fact, find a striking correlation
between $\th^{ph}$ and $\th^M$: The average ``photon'' angle $\th^{ph}$ tends 
to be positive for $\th^M \in [0,{\pi \over 2}]$, and negative for
$\th^M \in [{\pi \over 2},\pi]$.  Computing the average photon
angle $\overline{\th}^{ph}$ in each monopole angle quarter-interval,
we find
\beq
       \overline{\th}^{ph} = \left\{ \begin{array}{rl}
          0.027(4) & \mbox{for~~} \th^M \in [0,{\pi \over 2}] \cr
         -0.027(4) & \mbox{for~~} \th^M \in [{\pi \over 2},\pi]
         \end{array} \right. 
\label{oline}
\eeq
This result, combined with the results displayed in Table \ref{table1},
raises two interesting questions:
\begin{enumerate}
\item How is the correlation between $\tp$ and $\tm$, found above in
\rf{oline}, related to the
remaining $Z_2$ global symmetry of $\P(\th)$; and
\item Why is $P_1$ negative in the additive configurations of
eq.\ \rf{add}?
\end{enumerate}

  To shed some light on these issues, we begin by defining $\tpom$
as the average value of $\tp$ at fixed $\tm$,
and then make the drastic approximation of neglecting all fluctuations
of $\tp$ at fixed $\tm$ around its mean value.  This amounts to 
approximating the vev of any periodic function $F(\th)$
\beq
      <F> = \intpi d\th ~ F(\th) \P(\th)
\label{B1}
\eeq
by
\beq
      <F> = \intpi d\tm {1\over 2\pi} F(\tm + \tpom)
\label{B2}
\eeq
where we have used the fact that
the probability distribution for $\tm$ is (nearly) uniform.
The accuracy of this approximation depends, of course, on the width
of the probability distribution for $\tp$ at fixed $\tm$, and on
the particular $F(\th)$ considered.  Here we are only concerned
with certain qualitative aspects of phase angle probability distributions,
and hopefully the neglect of fluctuations of $\tp$ around the mean
will not severely mislead us.

   With the help of the approximation \rf{B2}, we can answer the
two questions posed above.  In this section we will only outline the argument,
which is presented in full in an Appendix.  

   The function $\tpom$ maps the variable $\tm \in [-\pi,\pi]$, 
which has a uniform probability distribution in the interval, into 
the variable $\to \in [-\pi,\pi]$, where
\beq
      \to = \tm + \tpom
\eeq
The non-uniform mapping induces a non-uniform probability distribution
for the $\to$-variable
\bea
      \P(\to) &=& {1\over 2\pi}{d\tm \over d\to} 
\non \\
              &=& {1\over 2\pi}\left(1 - {d\tpo \over d\to}\right)
\label{probo}
\eea
which we identify with $\P(\th)$ in the approximation \rf{B2}. 
Since $\P(\th)$ is peaked at $\th=\pm {\pi\over 2}$, it follows
that $d\tpo / d\to$ is minimized at $\to=\pm {\pi\over 2}$.

   Global $Z_2$ symmetry implies that $P_q=0$ for $q=$ odd.
Then, from eq. \rf{expand} we have
\bea
      \P(\pi-\to) &=& \P(\to)
\non \\
      \P(-\pi-\to) &=& \P(\to)
\non \\
      \P(-\to) &=& \P(\to)
\label{B16}
\eea
From these relationships, eq.\ \rf{probo}, and the fact (shown in
the Appendix) that $\tpo(-\tm)=-\tpom$, we find that 
\beq
      \tpo[\to] \equiv \tpo[\tm(\to)]
\eeq
is an odd function with respect to reflections around
$\to=0,\pm {\pi\over 2}$. Defining $\tpo_I$ as the average
$\tp$ in the quarter interval $\tm \in [0,{\pi\over 2}]$, and
$\tpo_{II}$ as the average $\tp$ in the quarter-interval
$\tm \in [{\pi\over 2},\pi]$, we have
\bea
     \tpo_I &=& {2\over \pi} \int_0^{\pi/2} d\tm ~ \tpom
\non \\
           &=& 4 \int_0^{\pi/2} d\to ~ \P(\to) \tpo[\to]
\non \\
     &=& 4 \int_{\pi/2}^{\pi} d\to ~ \P(\pi-\to) \tpo[\pi-\to]
\non \\
     &=& - 4 \int_{\pi/2}^{\pi} d\to ~ \P(\to) \tpo[\to] 
\non \\
     &=& - {2\over \pi} \int_{\pi/2}^{\pi} d\tm ~ \tpom 
\non \\
     &=& - \tpo_{II}
\label{B20}
\eea
which explains, as a consequence of global $Z_2$ symmetry, 
the equal magnitudes and opposite signs found in eq.\ \rf{oline}.
This answers the first of the two questions posed above.

\begin{figure}[h!]
\centerline{\scalebox{0.50}{\includegraphics{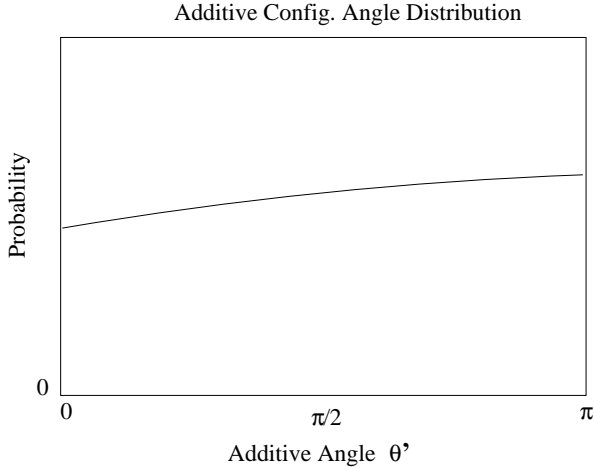}}}
\caption{Polyakov phase distribution $\P'(\th')$ in the
additive configuration $\th'=\th+\th^M$.}
\label{Padd}
\end{figure}

   For expectation values of Polyakov phase angles in the additive
configuration, we have
\beq
      <F> = \intpi d\th' ~ F(\th') \P'(\th')
\label{B3}
\eeq
where $\P'(\th')$ is the probability distribution for the
Polyakov angles of the additive configuration $\th'=\th+\tm$.
Again neglecting fluctuations of $\tp$ around the mean $\tpom$,
and changing variables to $\to'=\to + \tm$ we have
\bea
      <F> &=& \intpi d\tm ~ {1\over 2\pi} F(2\tm + \tpom)
\non \\
          &=& \int^{2\pi}_{2\pi} d\to' ~ {1\over 2\pi} {d\tm \over d\to'}
                      F(\to')
\non \\
          &=& \intpi d\to' ~ {1\over \pi} {d\tm \over d\to'}
                      F(\to')
\label{B4}
\eea
where the $2\pi$-periodicity of the integrand was used in the last step.
Then the induced probability distribution in the $\to'$ variable is
\bea
       \P(\to') &=& {1\over \pi} {d\tm \over d\to'}
\non \\
                &=& {1\over \pi} \left( 1 - {d\tpo \over d\to} \right)
                   \left( 2 - {d\tpo \over d\to} \right)^{-1}
\eea
As shown in the Appendix, $\to=\pm {\pi\over 2}$ corresponds to
$\to' = \pm \pi$, and the assumed single-valuedness
of $\tm(\to)$ requires ${d\tpo / d\to} < 1$.  In that case, since
${d\tpo / d\to}$ is minimized at $\to=\pm {\pi\over 2}$, it follows that
$\P(\to')$ has a peak at $\to'=\pm \pi$.
Given that the $\th'$ distribution is peaked at $\pm \pi$, as in
Fig.\ \ref{Padd}, the $n=1$ 
coefficient in the cosine series expansion of this distribution (which 
by definition is $P_1$) is evidently negative, answering the second of
the two questions posed below \rf{oline}.  Confinement is lost because the
$Z_2$ symmetry of the $\th$ distribution has been broken.

   The correlation that exists between the monopole and photon contributions
in abelian-projected SU(2) gauge theory implies that these contributions
actually do \emph{not} factorize in Polyakov lines and Wilson loops, in
contrast to the factorization which occurs in compact QED in the Villain
formulation.  In fact, the terminology ``photon contribution'' used to
describe $\tp$ is really a little misleading.  The field $\tp_\m(x)$ is
best described as simply the difference $\th_\m(x)-\tm_\m(x)$ between
the abelian angle field and the MD angle field.  It is not correct to
view $\tp_\m$ as a purely perturbative contribution, since the correlation
that exists between $\tp_\m$ and $\tm_\m$, which breaks U(1) down to
an exact $Z_2$ remnant symmetry, clearly has a non-perturbative origin.

\begin{figure}[h!]
\centerline{\scalebox{0.75}{\includegraphics{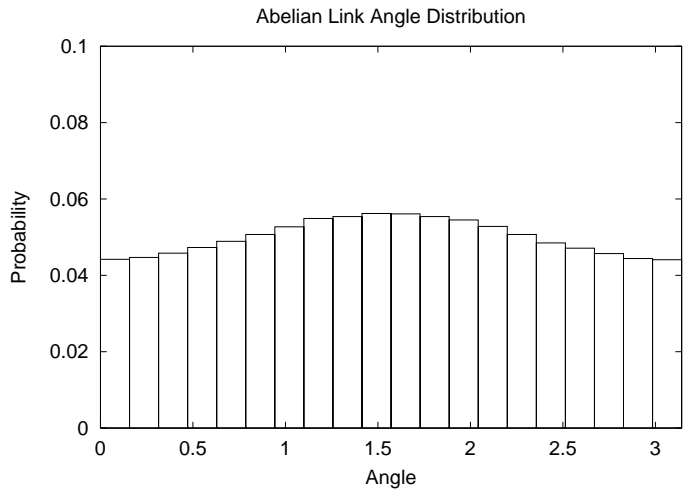}}}
\caption{Histogram of $\P(\th)$ in the interval $[0,\pi]$, 
for $T=3$ links in the time direction at $\b=2.1$. Note the
symmetry around $\th=\pi/2$, which is associated with $Z_2$
symmetry.}
\label{link_ang}
\end{figure}

\begin{figure}[h!]
\centerline{\scalebox{0.75}{\includegraphics{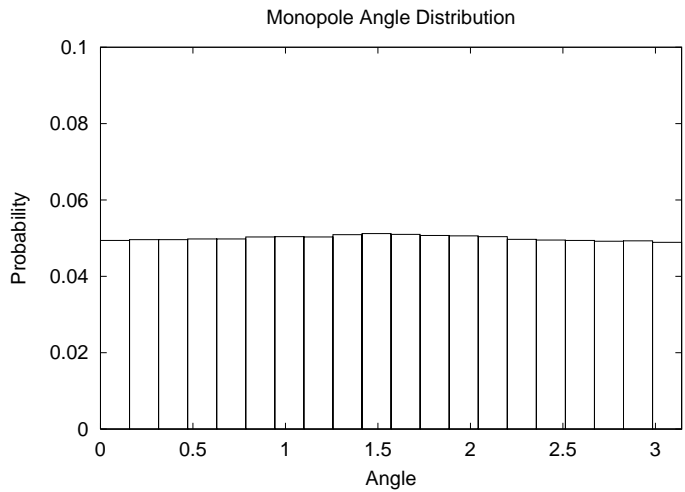}}}
\caption{Histogram of $\P(\th^M)$, for Polyakov lines in the
MD approximation.  Time extent is $T=3$ at $\b=2.1$.}
\label{mpole}
\end{figure}

\begin{figure}[h!]
\centerline{\scalebox{0.75}{\includegraphics{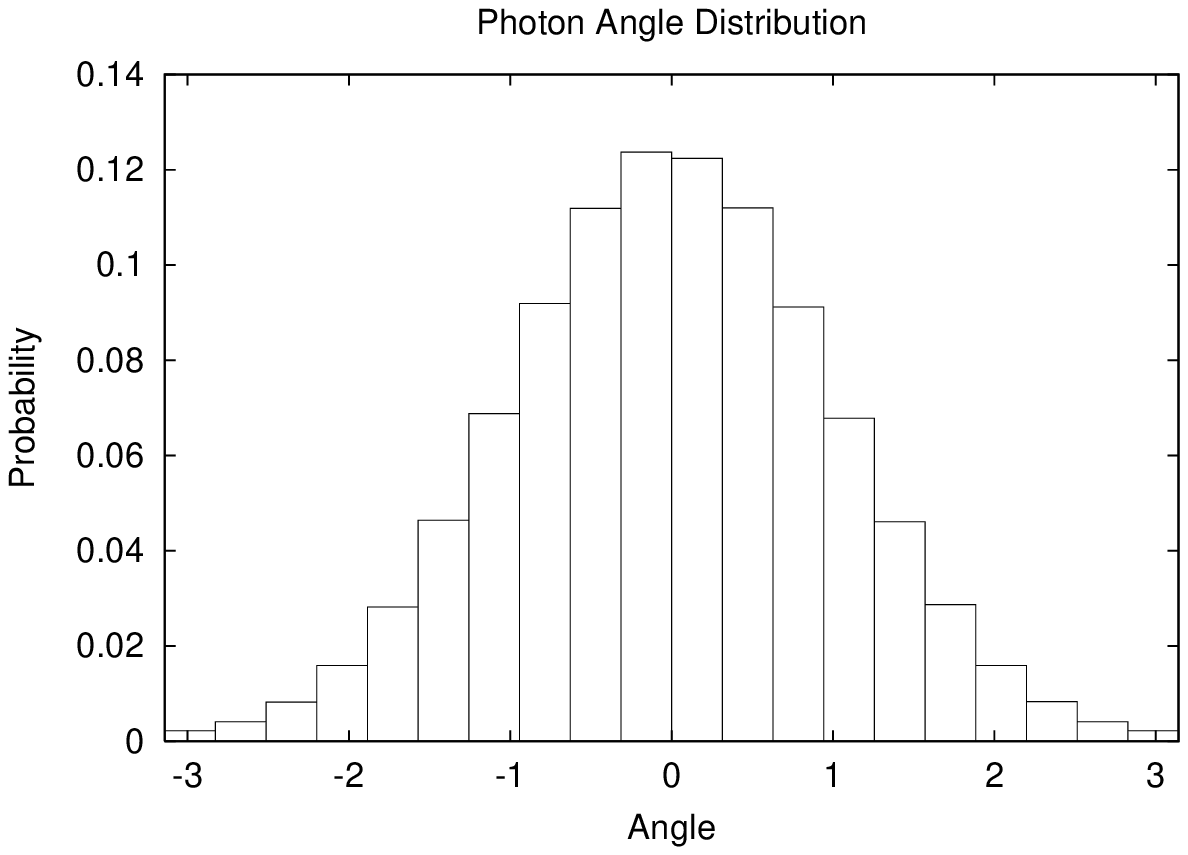}}}
\caption{Histogram of $\P(\th^{ph})$, for the ``photon'' Polyakov lines,
on the full interval $[-\pi,\pi]$. Time extent is $T=3$ at $\b=2.1$.}
\label{photon}
\end{figure}

    Finally, in Figs. \ref{link_ang}-\ref{photon}, we show some 
histograms for the actual probability
distributions $\P(\th),\P(\th^M),\P(\th^{ph})$, obtained on a 
$3\times 12^3$ lattice at $\b=2.1$.  $\P(\th)$ and $\P(\th^M)$ are
shown on the $[0,\pi]$ half-interval, while $\P(\th^{ph})$ is displayed
on the full $[-\pi,\pi]$ interval. The height of the histogram is the
probability for $|\th|,|\th^M|,\th^{ph}$ to fall in each interval.   
It is clear that these numerical results agree with the conjectured 
behavior in Figs.\ \ref{Ptheta}-\ref{Pphoton}.

\setcounter{equation}{0}
\section{Field Collimation}

   Although the finite VEV of $q=2$ Polyakov lines is a crucial
test, it is also useful to ask whether the collimation of confining field
strength into vortex tubes can be seen more directly on the lattice.

   In the most naive version of the monopole Coulomb gas, the monopole
field is imagined to be distributed symmetrically, modulo some small 
quantum fluctuations, around a static monopole.  In this section we will find
that the field around the position of an abelian monopole, as probed
by SU(2)-invariant Wilson loops, is in fact highly asymmetric, and is
very strongly correlated with the direction of the center vortex
passing through the monopole position.  Some of these results, for
unit cubes around monopoles, have been reported previously in ref.\
\cite{Zako}, but are included here for completeness.  The results for
3- and 4-cubes around static monopoles are new.  

   To circumvent the Gribov copy issue, we work in the ``indirect''
maximal center gauge introduced in ref. \cite{indirect}, and locate
monopole and vortex positions by projections (abelian and center,
respectively) of the same gauge-fixed configuration.  Indirect maximal
center gauge is a partial fixing of the U(1) gauge symmetry remaining 
in maximal abelian gauge, so as to maximize the squared trace of abelian 
links.  The residual gauge symmetry is $Z_2$.  The excitations of
the center projected lattice are termed ``P-vortices,'' and have
been found to lie near the middle of thick center vortices on the
unprojected lattice (cf. \cite{Jan98}).

\subsection{Monopole-Antimonopole Alternation}

\begin{figure}[h!]
\centerline{\scalebox{1.0}{\includegraphics{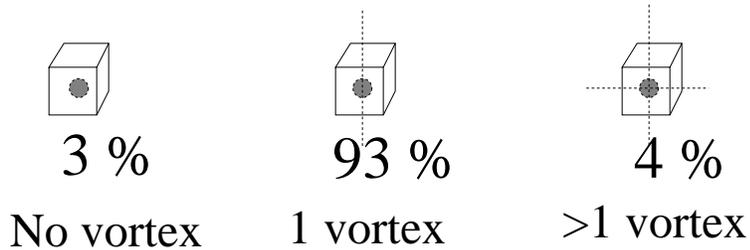}}}
\caption{Percentages of monopole cubes pierced by zero, one,
or more than one P-vortex lines, at $\b=2.4$.}
\label{acube1}
\end{figure}

   According to the argument depicted in Figs.\ 
\ref{avort1}-\ref{avort3}, at any fixed time the monopoles 
found in abelian projection should lie along vortex lines,
with monopoles alternating with antimonopoles along the line.
To test this argument, we consider static monopoles (associated with
timelike monopole currents) on each constant time volume of the
lattice.  Each monopole is associated with a net $\pm 2\pi$ magnetic flux
through a unit cube.  In numerical simulations performed at $\b=2.4$, we find
that almost every cube, associated with a static monopole, is pierced
by a single P-vortex line.  Only very small fractions are either not 
pierced at all, or are pierced by more than one line,
with percentages shown in Fig.\ \ref{acube1}.  

   P-vortices are line-like objects on any given time slice of
the lattice.\footnote{It should be noted that vortices are surface-like 
objects in D=4 dimensions, so different
closed loops on a given time slice may belong to the same P-vortex surface.}
About 61\% of these vortex lines have no monopoles at all on them.  We find
that 31\% contain a monopole-antimonopole pair.  The remaining 8\%
of closed vortex lines have an even number of monopoles + antimonopoles, 
{\it with monopoles
alternating with antimonopoles} as one traces a path along the loop. 
This is exactly the situation sketched in Fig. \ref{avort5}.
Exceptions to the monopole-antimonopole alternation rule were found in
only 1.2\% of loops containing monopoles.  In every exceptional
case, a monopole or antimonopole was found within one lattice unit of the 
P-vortex line which, if counted as lying along the vortex line, would
restore the alternation.

\subsection{Field Collimation on 1-cubes}

   We define vortex limited Wilson loops $W_n(C)$ as the expectation
value of Wilson loops on the full, unprojected lattice, subject to the
the constraint that, on the projected lattice, exactly $n$ P-vortices
pierce the minimal area of the loop (cf. \cite{Jan98}).  We employ
these gauge-invariant loop observables to probe the (a)symmetry
of the color field around static monopoles, again at $\b=2.4$.

\begin{figure}[h!]
\centerline{\scalebox{1.0}{\includegraphics{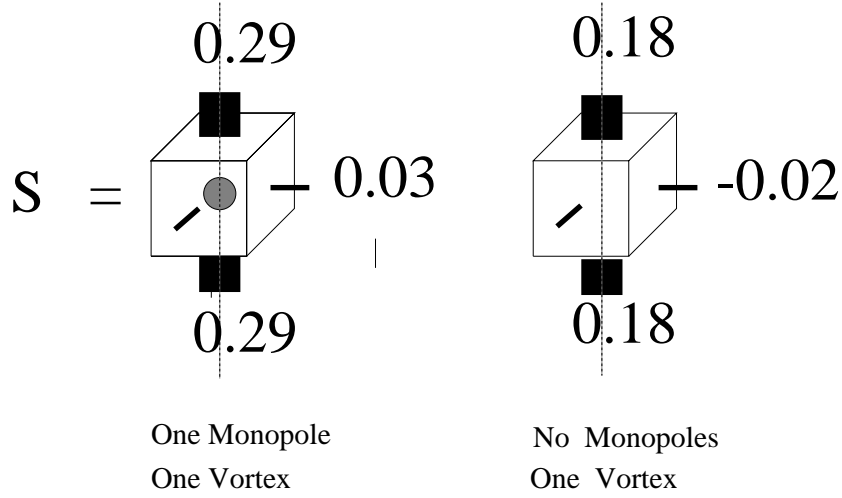}}}
\caption{Excess plaquette action distribution on
a monopole cube pierced by a single P-vortex. For comparison,
the excess action distribution is also shown for a no-monopole
cube pierced by a P-vortex.}
\label{acube3}
\end{figure}

   Consider first, on a fixed time hypersurface, the set of all 
unit cubes which contain one static monopole, inside a cube
pierced by a single P-vortex line.  This means that two plaquettes
on the cube are pierced by the vortex line, and four are not.
The difference $S$ between the average plaquette $S_0$ on the lattice,
and the plaquette on pierced/unpierced plaquettes of the monopole cube 
\beq
         S = S_0 - <\oh \mbox{Tr}[UUU^\dg U^\dg]_{\mbox{cube face}}>
\eeq
is shown in Fig.\ \ref{acube3}. For comparison, we have computed the 
same quantities in unit cubes, pierced by vortices, which do not contain 
any monopole current.

\begin{figure}[h!]
\centerline{\scalebox{1.0}{\includegraphics{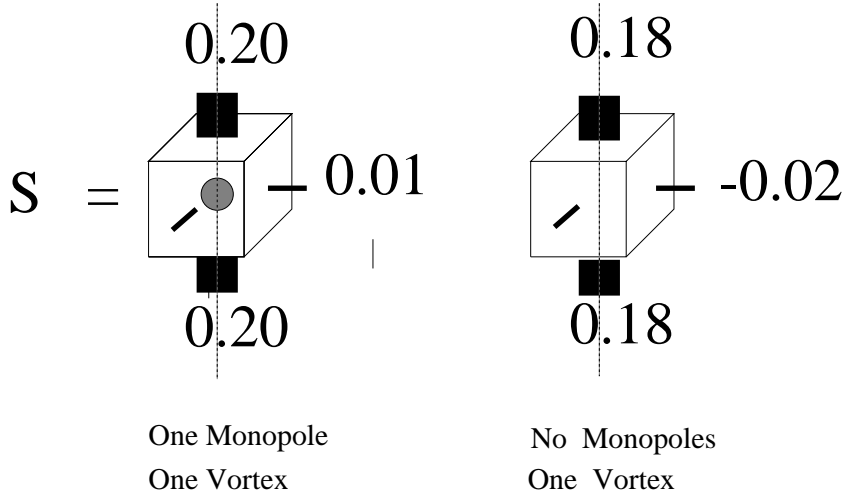}}}
\caption{Same as Fig.\ \ref{acube3} for isolated monopoles.}
\label{acube4}
\end{figure}

   It is obvious that the excess plaquette action associated with
a monopole is extremely asymmetric, and almost all of it is
concentrated in the P-vortex direction.  Moreover, the action distribution
around a monopole cube is not very different from the distribution
on a cube pierced by a vortex, with no monopole at all inside.
The two distributions are even more similar, if we make the
additional restriction to ``isolated'' static monopoles; i.e.\ monopoles
with no nearest-neighbor monopole currents.  The excess action distribution
for isolated monopoles, again compared to zero-monopole one-vortex cubes, is
shown in Fig.\ \ref{acube4}.

\subsection{Field Collimation on Larger Cubes}

   Finally we consider cubes which are
$N$ lattice spacings on a side, in fixed-time hypersurfaces, 
having two faces pierced by a single
P-vortex line and the other four faces unpierced.  Again we restrict 
our attention to cubes containing either
a single static monopole, or no monopole current.  Each side of the
cube is bounded by an $N\times N$ loop.  Let
\bea
   W_1^M(N,N) &\equiv& \mbox{1-vortex loops, bounding a monopole N-cube}
\non \\
   W_0^M(N,N) &\equiv& \mbox{0-vortex loops, bounding a monopole N-cube}
\non \\
   W_1^0(N,N) &\equiv& \mbox{1-vortex loops, bounding a 0-monopole N-cube}
\non \\
   W_0^0(N,N) &\equiv& \mbox{0-vortex loops, bounding a 0-monopole N-cube}
\eea
denote the expectation value of $N\times N$ Wilson loops on 0/1-vortex
faces of 0-monopole/1-monopole N-cubes.
As a probe of the distribution of gauge-invariant flux around an N-cube, we
compute the fractional deviation of these loops from 
$W_0^0(N,N)$ (which has the largest value) by
\bea
  A^M_{0,1} = {W^0_0(N,N) - W^M_{0,1}(N,N) \over W^0_0(N,N)}
\non \\
  A^0_{0,1} = {W^0_0(N,N) - W^0_{0,1}(N,N) \over W^0_0(N,N)}
\eea
and of course $A^0_0=0$ by definition.  

\begin{figure}[h!]
\centerline{\scalebox{1.0}{\includegraphics{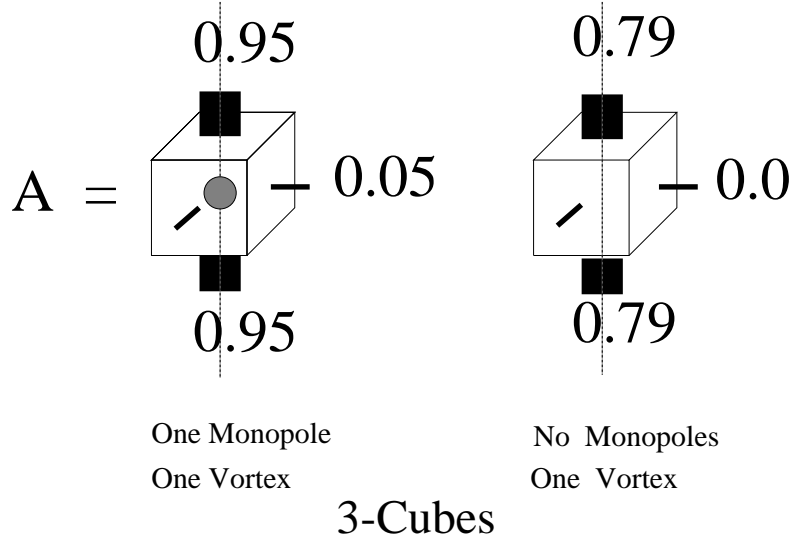}}}
\caption{Gauge field collimation on a 3-cube pierced by
a single P-vortex, for both one-monopole and no-monopole
3-cubes.  $A$ is the fractional deviation of $W(3,3)$ on a cube
face from the no-monopole, no-vortex value $W_0^0(3,3)$.}
\label{mcube3}
\end{figure}

\begin{figure}[h!]
\centerline{\scalebox{1.0}{\includegraphics{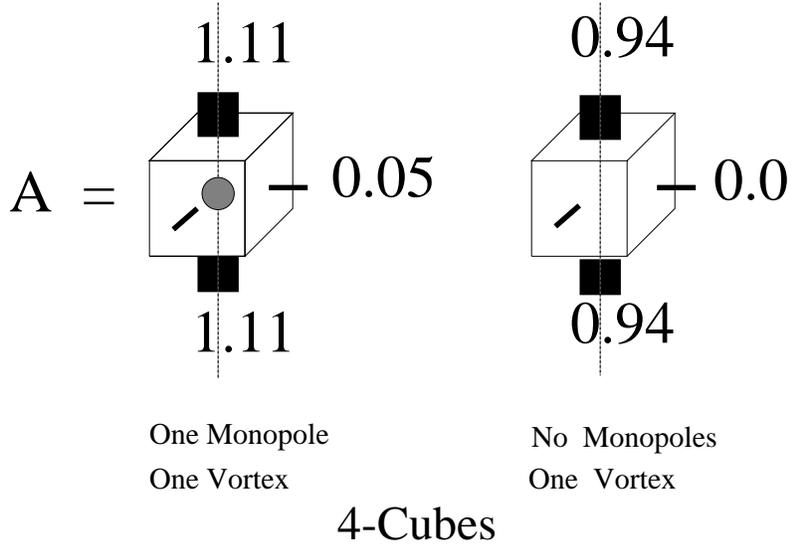}}}
\caption{Same as Fig.\ \ref{mcube3}, for 4-cubes.}
\label{mcube4}
\end{figure}

   The results, for $N=3$ cubes and $N=4$ cubes, are displayed in Figs.\
\ref{mcube3}-\ref{mcube4}, with the actual values for the various loop types
listed in Table \ref{table3}.  As with the excess-action distribution
around a 1-cube, shown in Figs.\ \ref{acube3}-\ref{acube4}, it is
clear that gauge-invariant Wilson loop values are distributed
very asymmetrically around a cube, and are strongly correlated 
with the direction of the P-vortex line.  The presence or absence
of a monopole inside the $N$-cube appears to have only a rather
weak effect on the value of the loops around each side of the cube;
the main variation is clearly due to the presence or absence of a vortex line
piercing the side. Obviously, the strong correlation of loop values
with vortex lines, and the relatively weak correlation of loop values
with monopole current, fits the general picture discussed in the 
introduction, of confining flux collimated into tubelike structures.

\begin{table}
\centerline{
\begin{tabular}{|l|r|r|r|r|} \hline\hline
(N)-cube & $W_1^M(N,N)$  & $W_0^M(N,N)$ &
           $W_1^0(N,N)$  & $W_0^0(N,N)$ \\ \hline
1-cube   &  0.3349 (10)  & 0.6007 (5)   &
            0.4340 (5)   & 0.6472 (2)  \\ 
2-cube   &  0.0599 (8)   & 0.2426 (5)   &
            0.1101 (6)   & 0.2587 (3)  \\ 
3-cube   &  0.0045 (7)   & 0.0825 (5)   &
            0.0179 (5)   & 0.0864 (4)  \\ 
4-cube   & -0.0030 (7)   & 0.0255 (7)   &
            0.0016 (6)   & 0.0268 (5)  \\ \hline
\end{tabular} }
\caption{Wilson loops on the faces of one-monopole ($W^M_{0,1}$) and
zero monopole ($W^0_{0,1}$) N-cubes. The $0,1$ subscript refers to
the number of P-vortices piercing the loop.}
\label{table3}
\end{table}

   It seems clear that the gauge field-strength distribution
around monopoles identified in maximal abelian gauge is highly
asymmetric, and closely correlated to P-vortex lines, as one would 
naively guess from the picture shown in Fig.\ 1.  This fact, by itself, 
doesn't prove the vortex theory.  One must admit the possibility
that if the monopole positions were somehow held fixed, fluctuations in
the direction of vortex lines passing through those monopoles might 
restore (on average) a Coulombic distribution.  We therefore regard
the findings in this section as fulfilling a necessary, rather than 
a sufficient, condition for the vortex theory to be correct.  
According to the vortex theory, confining fields are collimated along 
vortex lines, and this collimation should be visible on the lattice in 
some way, e.g.\ showing up in spatially asymmetric distributions of 
Wilson loop values.  This strong asymmetry is, in fact, what we find
on the lattice.

\setcounter{equation}{0}
\section{\boldmath The Seiberg-Witten confinement scenario and \\
$Z_2$-fluxes}\label{witten}

The beautiful description of confinement by monopoles when   
$N\!\!=\!2$ supersymmetric Yang-Mills theory is softly broken
to $N\!\!=\!1$ supersymmetric Yang-Mills theory has enhanced the 
impression that {\it the} mechanism for confinement in
pure Yang-Mills theory is monopole condensation.
Above we have provided evidence that it is in fact $Z_2$ fluctuations 
which are responsible for confinement in Yang-Mills theory even on
abelian-projected lattices; see refs.\ \cite{Jan98,dFE} for the
evidence on the full, unprojected lattices.  As regards the
$N\!\!=\!2$ supersymmetric Yang-Mills theory 
softly broken to $N\!\!=\!1$, 
we would simply argue that the approximate low-energy effective theory of 
Seiberg and Witten is not able describe all aspects of confinement at
sufficiently large distance scales.  In particular, the low-energy effective 
theory cannot explain the perimeter law of double-charged Wilson loops. 

   It may be useful here to make a distinction between the full effective
action, obtained by integrating out all massive fields, and the 
{\it ``low-energy''} effective action, which neglects all non-local or 
higher-derivative terms in the full effective action.  The Seiberg-Witten 
calculation was aimed at determining the low-energy effective action
of the softly broken $N\!\!=\!2$ theory.  However, in a confining theory,
non-local terms induced by massive fields can have important effects
at long distances. 

   The relation between the Seiberg-Witten 
theory and pure Yang-Mills theory in four dimensions has many 
similarities to the relation between the Georgi-Glashow model
and pure Yang-Mills theory in three dimensions.\footnote{This 
analogy has been  used to explain the measured (pseudo)-critical exponents
in four-dimensional lattice $U(1)$-theory from the properties of 
$N\!\!=\!2$ supersymmetric Yang-Mills theory and its possible symmetry 
breakings to $N\!\!=\!1$ and $N\!\!=\!0$ \cite{aes}.} 
In both theories the presence 
of a Higgs field is of utmost importance for the existence of 
monopoles. In the Georgi Glashow model it is the existence of 
a monopole condensate which
is responsible for the mass of the (dual) photon and for 
confinement of the smallest unit of electric $U(1)$ charge. However, 
as emphasized in \cite{GG3}, the effective low energy 
Coulomb gas picture of monopoles 
does not explain the fact that double-charged Wilson loops follow a 
perimeter law rather than an area law. The reason is that 
{\it in a confining theory there is not equivalence between low-energy 
and long-distance physics}. At sufficiently long distance it will 
always be energetically favorable to excite charged massive $W$-fields
and screen $q=$ even external charges, thus preventing a genuine 
string tension between such charges. At these large scales, a
description in terms of U(1)-disordering configurations (the monopole
Coulomb gas) breaks down, and a description in terms of $Z_2$ disorder
must take over.  As shown in \cite{GG3} the range of validity of the
monopole Coulomb gas picture decreases as the 
mass of the $W$-field decreases, and, in the limit of an unbroken 
$SU(2)$ symmetry, confinement can only be described adequately
in terms of  $Z_2$ fluctuations.

In the Seiberg-Witten theory, $N\!\!=\!2$ supersymmetry ensures that 
the Higgs vacuum is 
parameterized by an order parameter $u= \lla \tr \phi^2 \rra$
corresponding 
to the breaking of $SU(2)$ to $U(1)$.  For large values of $u$ we have 
a standard scenario: at energy scales  $\mu\gg\sqrt{u}$ all 
field theoretical degrees of freedom contribute to the $\b$-function,
which corresponds to the asymptotically free theory. For  energies
lower than $\sqrt{u}$ only the $U(1)$ part of the theory is effective.
In these considerations the dynamical confinement scale 
$$\L_{N=2}^4= \mu^4\exp(-8\pi^2/g(\mu)^2)$$ 
obtained by the one-loop perturbative calculation plays no role. 
The remarkable observation by Seiberg and Witten was that even 
when $u \leq \L^2_{N=2}$, where one would naively 
expect that non-Abelian dynamics was 
important, the system remains in  the $U(1)$ Coulomb phase due 
to supersymmetric cancellations of non-Abelian quantum fluctuations. 
As $u$ decreases the effective electric charge  associated
with the unbroken $U(1)$ part of $SU(2)$ increases, while the masses of
the solitonic excitations which are present in the theory, will
decrease.
Dictated by monodromy properties of the so-called prepotential of the 
effective low energy Lagrangian, the monopoles become massless at 
a point $u \sim \L^2_{N=2}$ where the effective electric charge
has an infrared Landau pole and diverges.
However, in the neighborhood of $u \sim \L^2_{N=2}$, 
this strongly coupled theory has an effective Lagrangian 
description as a weakly coupled theory when expressed  
in terms of dual variables, namely a monopole hyper-multiplet  
and a dual photon vector multiplet. The perturbative coupling constant
is now $g_D=4\pi/g$ and the point where monopoles condense corresponds
to $g_D=0$.

A remarkable observation of Seiberg and Witten is that 
the breaking of $N\!\!=\!2$ to $N\!\!=\!1$ supersymmetry 
by adding a  mass term  superpotential will generate 
a mass gap, originating from a condensation of the monopoles.
By the dual Meissner effect this theory confines the electric 
$U(1)$ charge at distances larger than the inverse $N\!\!=\!2$
symmetry breaking scale. In terms of the underlying microscopic theory
it is 
believed that the reduction of symmetry from $N\!\!=\!2$ to  $N\!\!=\!1$ 
allows excitations closer to generic non-supersymmetric 
``confinement excitations'', but that the soft breaking ensures
that the theory is still close enough to the $N\!\!=\!2$ to remain an
effective $U(1)$ theory.
Thus we see that in the Seiberg-Witten scenario we can, by introducing 
the mass term superpotential,
describe a $U(1)$ confining-deconfining transition 
from the $N\!\!=\!2$ Coulomb phase to the 
$N\!\!=\!1$ confining phase\footnote{The breaking down to 
$N\!=\!0$ has been analyzed in a number of 
papers \cite{EVA,ALV}, where soft breaking via 
spurion fields of  $N\!\!=\!1$ 
and $N\!\!=\!2$ supersymmetric gauge theories are discussed (see also 
\cite{SEE}). These models are somewhat closer to 
realistic models for $QCD$ confinement, but the conclusions are,
from our perspective, the 
same as for the original Seiberg-Witten model, so we will not 
discuss these models any further.}.

But precisely as for the Georgi-Glashow model, the monopole condensate 
picture for the $N=1$ confining theory is incomplete in the sense that
it cannot describe the  
obvious fact that double-charged Wilson loops 
will have a perimeter law rather than an area law. Clearly, $q=$ even 
external charges can be screened by the massive charged $W$-fields 
in the softly broken $N=2$ supersymmetric Yang-Mills theory, a fact which has 
profound implications for the large-scale structure
of confining fluctuations. But neither the
non-local effects of the $W$-fields, nor the
$W$-fields  themselves,
appear in the low energy effective action, a fact which 
illustrates once more that long distance physics is not captured
by the (local) low energy effective action in a confining theory.

\setcounter{equation}{0}
\section{Discussion}

   A point which was stressed both in ref.\ \cite{GG3} and in the
last section (see also \cite{Bali1}),
and which is surely relevant to the results reported here, is that
charged fields in a confining theory can have a profound
effect on the far-infrared structure of the theory, {\it even if those
fields are very massive.}  As an obvious example, consider integrating
out the quark fields in QCD, to obtain an effective pure gauge theory.
This effective pure gauge theory does not produce an asymptotic area law 
falloff for Wilson loops, which means that confining field configurations
are somehow suppressed at large distances.  A second example is the
Georgi-Glashow model in D=3 dimensions ($GG_3$),  
as discussed in ref.\ \cite{GG3}. In this case the W-bosons are massive,
and if their effects at large scales are simply ignored, then the model 
would be essentially equivalent to the theory of photons and monopoles, 
i.e. a monopole Coulomb gas, analyzed many years ago by Polyakov
\cite{Polyakov}.  In the monopole Coulomb gas, all multiples of the 
elementary electric charge are confined; but this is not what actually
happens in the Georgi-Glashow model.  The reason is that W-bosons are 
capable of screening even multiples of electric charge, which means that 
even-charge Wilson loops fall only with a perimeter law, and 
even-charge Polyakov lines have finite vacuum expectation values in the 
confined phase.  If we
again imagine integrating out the W and Higgs fields, then the effective
abelian theory confines only odd multiples of charge, the global
symmetry is $Z_2$, rather than U(1), and the theory is clearly not
equivalent to either a monopole Coulomb gas, or to compact $QED_3$.  
If one asks: how can the effective long-range theory, which involves
only the photon field, be anything different from compact $QED_3$,
the answer is that the integration over W and Higgs fields produces
non-local terms in the effective action.  We note, once again, that charged 
fields in
a confining theory have very long-range effects.  The fact that these
fields are massive does not imply that they can only lead, in the effective
abelian action, to local terms, or that the non-local terms can be neglected
at large scales.  These remarks also apply to
the Seiberg-Witten model, as discussed in the last section.

   In this article we have concentrated largely on a third example: 
abelian-projected Yang-Mills theory in maximal abelian gauge.  Calculations
on the abelian-projected lattice can be always regarded as being performed in
an effective abelian theory, obtained by integrating out the off-diagonal
gluon fields (and ghosts) in the given gauge.  It is often argued that the 
off-diagonal gluon fields are massive, and therefore do not greatly affect the 
long-range structure of the theory.  The long-range structure, according
to that view, is dominated exclusively by the diagonal gauge fields 
(the ``photons'') and the corresponding abelian monopoles,
which together are equivalent to a Coulomb gas of monopoles (D=3) or
monopole loops (D=4).  Then, since only abelian fields are involved, the
global symmetry of the effective long-range theory is expected to 
be U(1), and all multiples of abelian charge are confined.  
We have seen that reasoning of this sort, which neglects the long-distance 
effects of massive charged fields, can lead to erroneous conclusions.  In fact,
we have found that on the abelian projected lattice:
\begin{itemize}
\item Confinement of all multiples of abelian charge does 
\emph{not} occur on the abelian-projected lattice; charge $q=2$ 
Polyakov lines have a non-zero VEV.  
\item As a result, the global symmetry of the abelian-projected
lattice is at most $Z_2$, rather than U(1).
\item Monopole dominance breaks down rather decisively, at least when applied
to charge $q=2$ operators.
\item The distribution of Wilson loop values is highly asymmetric on
an N-cube.  There is a very strong correlation between loop values and
the P-vortex direction, but only a rather weak correlation with the
presence or absence of a static monopole in the N-cube.
\end{itemize}
In addition, in the usual maximal abelian gauge, there is a breakdown
of positivity, which is surely due to the absence of a transfer matrix
in this gauge.  The loss of positivity can be avoided (at the
cost of rotation invariance) by going to a spacelike maximal 
abelian gauge, where we again find $q=2$ string-breaking and
deconfinement.  The picture of a U(1)-symmetric monopole Coulomb gas
or dual superconductor,
confining all multiples of the elementary abelian charge, is clearly
not an adequate description of the abelian-projected theory at large
distance scales.  On the other hand, the results reported here fit
quite naturally into the vortex picture, where confining magnetic
flux on the projected lattice is collimated in units of $\pm \pi$.

   The center vortex theory has a number of well-known (and
gauge-invariant) virtues.  In particular, the vortex mechanism
is the natural way to understand, in terms of vacuum gauge-field
configurations, the screening of color charges in zero N-ality representations,
as well the loss of $Z_N$ global symmetry in the deconfinement phase 
transition \cite{Lang}.  Center vortex structure is visible on
unprojected lattices, through the correlation of P-vortex location with
gauge-invariant observables \cite{Jan98,dFE}.  The evidence we have
reported here, indicating vortex structure on large scales  
even on abelian-projected lattices, increases our confidence that center 
vortices are essential to the mechanism of quark confinement.

\vspace{32pt}

\ni {\Large \bf Acknowledgements}

\bigskip

  J.Gr.\ is happy to acknowledge the hospitality of the theory
group at Lawrence Berkeley National Laboratory, where some of this
work was carried out.  J.Gr.'s research is supported in part by  
the U.S.\ Department of Energy under Grant No.\ DE-FG03-92ER40711.

\appendix

\setcounter{equation}{0}
\section{Appendix}

\bigskip

   In this Appendix we present the detailed argument, outlined
in section 3, that $\tpo_I=-\tpo_{II}$ and $P_1<0$ in the additive
$\th' = \th + \tm$ configurations.  

   The approximation used here is to ignore,
at fixed $\th_M$, the fluctuctions of $\tp$ around the mean value
$\tpom$; i.e. the vev of any periodic function $F(\th)$ of the Polyakov
phase $\th$
\beq
      <F> = \intpi d\th ~ F(\th) \P(\th)
\label{A1}
\eeq
is approximated by
\beq
      <F> = \intpi d\tm {1\over 2\pi} F(\tm + \tpom)
\label{A2}
\eeq
where the factor of $1/2\pi$ corresponds to the uniform probability
distribution for $\tm$.  The mean value $\tpom$ is defined as
\bea
  \tpom &=& {1\over Z_{\tm}} \int D\th_\m(x) ~ \arg\left(
    P_1(\vx)e^{-i\tm}\right) \d \left[P_{M1}(\vx),e^{i\tm} \right]
    e^{-S_{eff}}
\non \\
  Z_{\tm} &=& \int D\th_\m(x) ~ \d \left[P_{M1}(\vx),e^{i\tm} \right]
    e^{-S_{eff}}    
\label{A3}
\eea
where
\bea
     P_1(\vx) &=& \prod_{n=1}^{N_T} \exp[i\th_4(\vx+n\hat{4})]
\non \\
     P_{M1}(\vx) &=& \prod_{n=1}^{N_T} \exp[i\tm_4(\vx+n\hat{4})]
\label{A4}
\eea
are Polyakov lines in the abelian and MD lattices, and $S_{eff}$ is
the effective abelian action, obtained after integrating out all
off-diagonal gluons and ghost fields.  Due to translation invariance,
$\tpom$ does not depend on the particular spatial position $\vx$ chosen
in \rf{A3}.

   From its definition, $\tpom$ is obviously periodic w.r.t.\ 
$\tm \ra \tm+2\pi$.  It is also an odd function of $\tm$, i.e.
\beq
        \tpom = - \tpo(-\tm)
\label{A5}
\eeq
This is derived by first noting that the $\tm_\mu(x)$ link angles
are functions of the $\th_\m(x)$ link angles according to eqs.\
\rf{MD1}-\rf{MD3}, and that $\tm_\m(x) \ra -\tm_\m(x)$ 
under the transformation 
$\th_\m(x) \ra -\th_\m(x)$.  Then, making the change of variables 
$\th_\m(x) \ra -\th_\m(x)$ in the integral \rf{A3}, we have
\bea
  Z_{\tm} &=& \int D\th_\m(x) ~ \d \left[P^*_{M1}(\vx),e^{i\tm} \right]
    e^{-S_{eff}}
\non \\
   &=& \int D\th_\m(x) ~ \d \left[P_{M1}(x),e^{-i\tm} \right]
              e^{-S_{eff}}
\non \\
   &=& Z_{-\tm}
\label{A6}
\eea
and
\bea
  \tpom &=& {1\over Z_\tm} \int D\th_\m(x) \mbox{arg}\left(
    P^*_1(x)e^{-i\tm}\right) \d \left[ P^*_{M1}(x),e^{i\tm} \right]
    e^{-S_{eff}}
\non \\
    &=& {1\over Z_{-\tm}} \int D\th_\m(x) (-1)\times \mbox{arg}\left(
    P_1(x)e^{i\tm}\right) \d \left[P_{M1}(x),e^{-i\tm} \right]
    e^{-S_{eff}}
\non \\
    &=& - \tpo(-\tm)
\label{A7}
\eea
The fact that $\tpom$ is an odd function of $\tm$, combined with
$2\pi$-periodicity, gives us
\beq
      \tpo(\pi) = \tpo(-\pi) = \tpo(0) = 0 
\label{A8}
\eeq

   We now define the variable
\beq
      \to(\tm) \equiv \tm + \tpom
\label{A9}
\eeq
which is the average Polyakov phase at fixed $\tm$.  It will be assumed
that $\to(\tm)$ is a single-valued function of $\tm$.
Eq.\ \rf{A9} can then be inverted to define $\tm$ implicitly as a
function of $\to$
\beq
      \tm(\to) = \to - \tpo[\tm(\to)]
\label{A11}
\eeq
and it will be convenient to introduce the notation
\beq
       \tpo[\to] \equiv \tpo[\tm(\to)]
\label{A12}
\eeq
Applying the change of variable \rf{A9} to eq.\ \rf{A2}, we have
\beq
  <F> = \intpi d\to ~ {1\over 2\pi} {d\tm \over d\to} F(\to)
\label{A13}
\eeq
where, from eqs.\ \rf{A8} and \rf{A9}, we see that the limits of integrations
are unchanged.  Comparing \rf{A13} to \rf{A1}, the
Polyakov phase probability distribution $\P(\th)$ can be identified
with
\bea
     \P(\to) &=&  {1\over 2\pi} {d\tm \over d\to}
\non \\
             &=&  {1\over 2\pi}\left( 1 - {d\tpo \over d\to} \right)
\label{A14}
\eea
in the approximation \rf{A2}.

   Assuming $Z_2$ symmetry in the confined phase, we have from eq.\
\rf{expand}
\beq
     \P(\to) = {1\over 2\pi}\left(1 + 
2 \sum_{q=\mbox{even}} P_q \cos(q\to) \right) 
\label{A15}
\eeq
which means that $\P(\to)$ is even w.r.t.\ reflections around
$\to=0,\pm {\pi \over 2}$; i.e.
\bea
      \P(\pi-\to) &=& \P(\to)
\non \\
      \P(-\pi-\to) &=& \P(\to)
\non \\
      \P(-\to) &=& \P(\to)
\label{A16}
\eea
Comparing eq.\ \rf{A16} with \rf{A14}, we find that ${d\tpo/d\to}$
is also even under reflections around $\to=0,\pm {\pi \over 2}$.  Since the
derivative of an odd function is an even function, this means that
\beq
     \tpo[\to] = a + \phi(\to)
\label{A17}
\eeq
where $\phi(\to)$ is odd under reflections around $\to=0$, and
$a$ is a constant.  However, since
\beq
      \to(\tm=0) = \tpo(0) = 0
\label{A18}
\eeq
it follows that $\tpo[\to=0]=\tpo(\tm=0)=0$.  Then $a=0$, and
$\tpo[\to]$ is odd around $\to=0$.  Further, from \rf{A8}, \rf{A9},
and the assumed single-valuedness of $\to(\tm)$, it follows that
$\tm(\to=\pm \pi) = \pm \pi$, and therefore that
\beq
         \tpo[\pm \pi] = 0
\eeq
Then, since $\tpo[0]=\tpo[\pi]=0$, and ${d\tpo/d\to}$ is even w.r.t.\
reflections around ${\pi \over 2}$, it follows that 
$\tpo[{\pi \over 2}]=0$, and that $\tpo[\to]$ is odd w.r.t.\ reflections
around ${\pi \over 2}$.  By the same reasoning, $\tpo[\to]$ is also odd
w.r.t.\ reflections around $-{\pi \over 2}$.  To summarize, $\tpo[\to]$
has the reflection properties:
\bea
       \tpo[-\to] &=& - \tpo[\to]
\non \\
       \tpo[\pi-\to] &=& -\tpo[\to]
\non \\
       \tpo[-\pi-\to] &=& -\tpo[\to]
\label{A19}
\eea
where the last two relationships are a consequence of 
global $Z_2$ symmetry in the confined phase.
Therefore
\bea
     \tpo_I &=& {2\over \pi} \int_0^{\pi/2} d\tm ~ \tpom
\non \\
            &=& 4 \int_0^{\pi/2} d\to ~ \P(\to) \tpo[\to]
\non \\
     &=& 4 \int_{\pi/2}^{\pi} d\to ~ \P(\pi-\to) \tpo[\pi-\to]
\non \\
     &=& - 4 \int_{\pi/2}^{\pi} d\to ~ \P(\to) \tpo[\to] 
\non \\
     &=& - {2\over \pi} \int_{\pi/2}^{\pi} d\tm ~ \tpom 
\non \\
     &=& - \tpo_{II}
\label{A20}
\eea
This explains why the correlation between $\tp$ and $\tm$ found numerically
in eq.\ \rf{oline} is a consequence of global $Z_2$ invariance.

   Our second task is to understand why $P_1$ is negative in the
$\th'=\th+\tm$ additive configuration, given that $\P(\th)$ is peaked
around $\th={\pi \over 2}$ as discussed in section 3.  Introducing the
probability distribution $\P'(\th')$ for the Polyakov phases in the
additive configurations
\beq
      <F(\to')> = \intpi d\to' ~ F(\to') \P'(\to')
\label{A21}
\eeq
and again neglecting the fluctuations of $\tp$ at fixed $\tm$ 
around the mean $\tpom$, 
\beq
      <F> = \intpi d\tm {1\over 2\pi} F(2\tm + \tpom)
\label{A22}
\eeq
Under the change of variables 
\beq
      \to' = 2\tm + \tpom
\label{A23}
\eeq
eq.\ \rf{A22} becomes
\beq  
      <F> = \int^{2\pi}_{-2\pi} d\to' {1\over 2\pi} {d\tm \over d\to'} F(\to')
\label{A24}
\eeq
The limits of integration have changed, but the original limits can
be restored using the $2\pi$-periodicity of the integrand.  To demonstrate
the periodicity, we first have
\bea
   \to'(\to+\pi) &=& \to + \pi + \tm(\to+\pi)
\non \\
    &=& \to + \pi + (\to + \pi - \tpo[\to+\pi])
\non \\
    &=& \to + \pi + (\to + \pi + \tpo[-\to])      
\non \\
    &=& \to + \pi + (\to + \pi - \tpo[\to])      
\non \\
    &=& \to'(\to) + 2\pi
\label{A25}
\eea
where the reflection properties \rf{A19} have been used.  Single-valuedness
of $\to'(\to)$ then implies the converse property
\beq
       \to(\to'+2\pi) = \to(\to') + \pi
\label{A26}
\eeq
Next,
\bea
    {d\tm \over d\to'} &=& {d\tm \over d\to}{d\to \over d\to'}
\non \\
                       &=& \left( 1 - {d\tpo \over d\to} \right)
                   \left( 2 - {d\tpo \over d\to} \right)^{-1}
\label{A27}
\eea
Then, applying \rf{A26} plus the fact that $d\tpo / d\to$ is even
w.r.t.\ reflections $\to \ra -\to$ and $\to \ra \pi - \to$,
\bea
    \left({d\tm \over d\to'}\right)_{\to' + 2\pi} &=& 
       \left( 1 - {d\tpo \over d\to} \right)_{\to(\to')+\pi}
       \left( 2 - {d\tpo \over d\to} \right)_{\to(\to')+\pi}^{-1}
\non \\
    &=&   \left( 1 - {d\tpo \over d\to} \right)_{-\to(\to')}
          \left( 2 - {d\tpo \over d\to} \right)_{-\to(\to')}^{-1}
\non \\
    &=&   \left( 1 - {d\tpo \over d\to} \right)_{\to(\to')}
          \left( 2 - {d\tpo \over d\to} \right)_{\to(\to')}^{-1}
\label{A28}
\eea
Since $F(\to')$ is periodic by definition, this establishes the
$2\pi$-periodicity of the integrand in \rf{A24}, which can then
be written
\bea  
      <F> &=& \int^{\pi}_{-\pi} d\to' {1\over \pi} {d\tm \over d\to'} F(\to')
\non \\
          &=&   \int^{\pi}_{-\pi} d\to' {1\over \pi}
                  \left( 1 - {d\tpo \over d\to} \right)
                   \left( 2 - {d\tpo \over d\to} \right)^{-1} F(\to')
\label{A29}
\eea
Comparing \rf{A29} with \rf{A21}
\beq
     \P(\to') = {1\over \pi} \left( 1 - {d\tpo \over d\to} \right)
                   \left( 2 - {d\tpo \over d\to} \right)^{-1}
\label{A30}
\eeq

   Single-valuedness of $\tm(\to)$ implies that $d\tpo / d\to < 1$,
and with this restriction $\P(\to')$ is a maximum where $d\tpo / d\to$ 
is a minimum.  However, we have previously deduced from the the fact that
$P_1=0$ and $P_2<0$ that the probability distribution
\beq
     \P(\to) = 1 - {d\tpo \over d\to}
\label{A31}
\eeq
is $Z_2$ invariant and peaked at $\to=\pm \pi/2$.  Again, this
distribution is maximized when $d\tpo / d\to$ is minimized, which
implies that $d\tpo / d\to$ is a minimum at $\to=\pm \pi/2$.
Finally,
\bea
     \to'\left(\to=\pm{\pi\over 2}\right) &=&
          {\pi\over 2} + \tm\left(\to={\pi\over 2}\right)
\non \\
       &=&  {\pi\over 2} + \left({\pi\over 2} - 
                    \tpo[{\pi\over 2}] \right)
\non \\
       &=& \pi
\label{A32}
\eea
As a consequence, $d\tpo / d\to$ is minimized at $\to' = \pi$
(and also, by the same arguments, at $\to'=-\pi$), which means that
$\P(\to')$ is peaked at $\to'=\pm \pi$, as illustrated in Fig.\ \ref{Padd}.
This explains why we expect $P_1<0$ in the additive configuration.

\end{document}